\documentclass[11pt]{article}
\begin{document}
\setlength{\oddsidemargin}{0cm} \setlength{\evensidemargin}{0cm}
\baselineskip=20pt

\begin{center} {\Large \bf Yangian and Applications}  \end{center}

\bigskip

\begin{center} { \large Cheng-Ming Bai, Mo-Lin Ge}\end{center}

\begin{center} {\it Theoretical Physics Division, Chern Institute of
Mathematics, Nankai University\\ Tianjin 300071, P.R.
China}\end{center}

\begin{center} {\large Kang Xue, Hong-Biao Zhang}\end{center}

\begin{center} {\it Department of Physics, Northeast Normal University, Changchun 130024, P.R.
China}\end{center}

\vspace{0.3cm}

\begin{center} {\large\bf   Abstract } \end{center}

In this paper, the Yangian relations are tremendously simplified
for Yangians associated to $SU(2)$, $SU(3)$, $SO(5)$ and $SO(6)$
based on RTT relations that much benefit the realization of
Yangian in physics. The physical meaning and some applications of
Yangian have been shown.

\bigskip

\section{Introduction}

Yangian was presented by Drinfel'd ([1-3]) twenty years ago. It
receives more attention for the following reasons. It is related
to the rational solution of Yang-Baxter equation and the RTT
relation. It is a simple extension of Lie algebras and the
representation theory of $Y(SU(2))$ has been given. Some physical
models, say, two component nonlinear Schrodinger equation,
Haldane-Shastry model and 1-dimensional Hubbard chain do have
Yangian symmetry. Yangian may be viewed as the consequence of a
``bi-spin" system. How to understand the physical meaning of
Yangian is an interesting topic. In this paper, there is nothing
with mathematics. Rather, we try to use the language of quantum
mechanics and Lie algebraic knowledge to show the effects of
Yangian.

\bigskip

\section{Yangian and RTT Relations}

Let ${\mathcal G}$ be a complex simple Lie algebra. The Yangian
algebra $Y({\mathcal G})$ associated to ${\mathcal G}$ was given
as follows ([1-3]). For a given set of Lie algebraic generators
$I_\mu$ of ${\mathcal G}$ the new generators $J_\nu$ were
introduced to satisfy
$$[I_\lambda,I_\mu]=C_{\lambda\mu\nu}I_\nu,\;\;C_{\lambda\mu\nu}\;\;{\rm are}\;\;{\rm structural}\;\;{\rm constants};
\eqno(2.0.1)$$
$$[I_\lambda,J_\mu]=C_{\lambda\mu\nu}J_\nu; \eqno(2.0.2)$$
and, for ${\mathcal G}\ne sl(2)$:
$$[J_\lambda,[J_\mu,I_\nu]]-[I_\lambda,[J_\mu,J_\nu]]=a_{\lambda\mu\nu\alpha
\beta\gamma}\{I_\alpha,I_\beta,I_\gamma\}, \eqno(2.0.3)$$ where
$$a_{\lambda\mu\nu\alpha\beta\gamma}=\frac{1}{4!}C_{\lambda\alpha\sigma}
C_{\mu\beta\tau}C_{ \nu\gamma\rho}C_{\sigma\tau\rho},
\eqno(2.0.4)$$
$$\{x_1,x_2,x_3\}=\sum_{i\neq j\neq k}x_ix_jx_k, \ \ ({\rm symmetric\ \ summation});\eqno(2.0.5)$$
or for ${\mathcal G}=sl(2)$:
$$[[J_\lambda,J_\mu],[I_\sigma,J_\tau]]+[[J_\sigma,J_\tau],[I_\lambda,J_\mu]]$$
$$=(a_{\lambda\mu\nu\alpha\beta\gamma}C_{\sigma\tau\nu}+
a_{\sigma\tau\nu\alpha\beta\gamma}C_{\lambda\mu\nu})\{I_\alpha,I_\beta,
J_\gamma\}.\eqno(2.0.6)$$

When
$C_{\lambda\mu\nu}=i\varepsilon_{\lambda\mu\nu}(\lambda,\mu,\nu=1,2,3)$,
equation (2.0.3) is identically satisfied from the Jacobian
identities. Besides the commutation relations there are
co-products as follows.
$$\Delta(I_\lambda)=I\lambda\otimes 1+1\otimes I_\lambda;\eqno
(2.0.7)$$
$$\Delta(J_\lambda)=J_\lambda\otimes 1+1\otimes
J_\lambda+\frac{1}{2}C_{\lambda\mu\nu}I_\mu\otimes I_\nu.\eqno
(2.0.8)$$

Further, the Yangian can be derived through RTT relations where
$R$ is a  rational solution of Yang-Baxter equation (YBE)
([1-12]).

After lengthy calculations, we found the independent relations for
$Y(SU(2))$, $Y(SU(3))$, $Y(SO(5))$ and $Y(SO(6))$ by expanding the
RTT relations and also checked through equations (2.0.1)-(2.0.3)
and (2.0.6) by substituting the structural constants ([13-17]),
where RTT relation (Faddeev, Reshetikhin, Takhtajan --- RFT [18])
satisfies
$$\check{R}(u-v)(T(u)\otimes 1)(1\otimes T(v))=(1\otimes
T(v))(T(u)\otimes 1){\check{R}}(u-v). \eqno (2.0.9)$$

\bigskip

\subsection{$Y(SU(2))$}

Let $P_{12}$ be the permutation. Setting
$$
{\check{R}}_{12}(u)=PR_{12}(u)=uP_{12}+I;\eqno (2.1.1)$$
\begin{eqnarray*}\hspace{2cm}T(u)&=&I+\sum^\infty_{n=1}u^{-n} {\left [ \matrix{
T^{(n)}_{11} & T_{12}^{(n)} \cr T_{21}^{(n)} & T^{(n)}_{22}
}\right ]}\\ &=&I+\sum^\infty_{n=1}u^{-n}
{\left[\matrix{\frac{1}{2}(T^{(n)}_0+T^{(n)}_3), & T^{(n)}_+ \cr
T^{(n)}_-, &
\frac{1}{2}(T^{(n)}_0-T^{(n)}_3)}\right]},\hspace{2.6cm} (2.1.2)
\end{eqnarray*}
and substituting the $T(u)$ into RTT relation it turns out that
only
$$I_{\pm}=T^{(1)}_{\pm}, I_3=\frac{1}{2}T^{(1)}_3;\eqno (2.1.3)$$
$$J_\pm=T_\pm^{(2)},  J_3=\frac{1}{2}T_3^{(2)}\eqno (2.1.4)$$
are independent ones. The quantum determinant
$$\det
T(u)=T_{11}(u)T_{22}(u-1)-T_{12}(u)T_{21}(u-1)=C_0+\sum^\infty_{n=1}u^{-n}C_n\eqno
(2.1.5)$$ gives
$$ C_0=1,\ \ C_1=T^{(1)}_0=trT^{(1)},\eqno (2.1.6)$$
$$C_2=T^{(2)}_0-{\bf I}^2+T^{(1)}_0(1+\frac{1}{2}T^{(1)}_0),\;\;\cdots,\;.\eqno (2.1.7)$$
The independent commutation relations of $Y(SU(2))$ are:
$$[I_\lambda,I_\mu]=i\epsilon_{\lambda\mu\nu}I_\nu\ \ \ \ \
(\lambda,\mu,\nu=1,2,3);\eqno (2.1.8)$$
$$[I_\lambda,J_\mu]=i\epsilon_{\lambda\mu\nu}J_\nu;\eqno (2.1.9)$$
and $(A_\pm=A_1\pm iA_2)$
$$[J_3,[J_+,J_-]]=(J_-J_+-I_-J_+)I_3\eqno (2.1.10)$$
that can be checked to generate all of relations of equations
(2.0.1), (2.0.2) and (2.0.6) with the help of Jacobi identities.

The co-product is given through (RFT) as $$\Delta T_{ab}=\sum_c
T_{ac}\otimes T_{cb}.\eqno (2.1.11)$$

 \vskip 0.5cm

The simplest realization of $Y(SU(2))$ is
$${\bf I}=\sum^N_{i=1}{\bf I}_i\;\;  (i:\mbox{lattice
indices}),\eqno (2.1.12)$$ $${\bf J}=\sum^N_{i=1}\mu_i{\bf
I}_i+\sum^N_{i<j}W_{ij}{\bf I}_i\times{\bf I}_j,\eqno (2.1.13)$$
where
$$
 W_{ij}= \left\{ \matrix{1 & i<j\cr 0 & i=j \cr -1
& i>j}\right. \;\;({\rm for}\;\;{\rm any}\;\;{\rm
representation}\;\;{\rm of}\;\; SU(2)) \eqno (2.1.14)$$ or
$$W_{jk}= i\cot\frac{(j-k)\pi}{N}\;\;({\rm only}\; {\rm for}\;
{\rm spin}\;\frac{1}{2},\;{\rm Haldane}-{\rm Shastry}\;{\rm
model}\; [19-21]),\eqno(2.1.15)$$ and $\mu_i$ arbitrary constants.
Noting
 that $\mu_i$ plays important role for the representation theory of
 $Y(SU(2))$ given by Chari and Pressley ([22-24]).

 \vskip 0.5cm
 The big difference between representations of Lie algebra and
 Yangian is in that in Yangian there appear free parameters
 $\mu_i$ depending on models.

Another example for single particle is finite $W$-algebra
([25-26]). Denoting by ${\bf L}$ and ${\bf B}$ angular momentum
and Lorentz boost, respectively, as well as $D$ the dilatation
operator, the set of ${\bf L}$ and ${\bf J}$ satisfies $Y(SU(2))$
where ([13],[25])
$${\bf I}={\bf L}\eqno (2.1.16)$$
$${\bf J}={\bf I}\times{\bf B}-i(D-1){\bf B}\eqno (2.1.17)$$
and $$[J_\alpha,J_\beta]=i\epsilon_{\alpha\beta\gamma}(2{\bf
I}^2-c_2^\prime-4){\bf I}_\gamma,\;\;c_2^\prime\ \ \mbox{casimir
of}\ \ SO(4,2).\eqno (2.1.18)$$

\vskip 0.5cm

 There are the following models whose Hamiltonians do commute with $Y(SU(2))$.

 $\bullet$ Two component nonlinear Schrodinger equation (Murakami and Wadati [27])
$$i\psi_t=-\psi_{xx}+2c|\psi|^2\psi,\eqno (2.1.19)$$
$${\bf I}=\int dx \psi_\alpha^+(x)(\frac{\sigma}{2})_{\alpha\beta}\psi_\beta(x);\eqno (2.1.20)$$
$${\bf J}=-i\int dx
\psi_\alpha^+(x)(\frac{\sigma}{2})_{\alpha\beta}\psi_\beta(x)-\frac{ic}{2}\int
dxdy\varepsilon
(y-x)(\frac{\sigma}{2})_{\beta\lambda}\psi_\beta^+(x)\psi_\alpha^+(y)
\psi_\alpha(x)\psi_\lambda(y).\eqno (2.1.21)$$

 $\bullet$ One-dimensional Hubbard model (for $N\rightarrow \infty$, [28])
$$H=-\sum_{i=1}^N(a_i^+a_{i+1}+a^+_{i+1}a_i+b_i^+b_{i+1}+b_{i+1}^+b_i)-U\sum_{i=1}^N
(a_i^+a_i-\frac{1}{2})(a_i^+a_i-\frac{1}{2}); \eqno (2.1.22)$$
\begin{eqnarray*}
\hspace{3cm}J_\pm&=&J_1\pm iJ_2,\\
J_+ &=& \sum_{i,j}\theta_{i,j}a_i^+b_j-U\sum_{i\ne
j}\varepsilon_{i,j}I_i^+I_j^3,\\
J_- &=& \sum_{i,j}\theta_{i,j}b_i^+a_j+U\sum_{i\ne
j}\varepsilon_{i,j}I_i^-I_j^3,\\
J_3 &=& \frac{1}{2}[
\sum_{i,j}\theta_{i,j}(a_i^+a_j-b_i^+b_j)+U\sum_{i<
j}\varepsilon_{i,j}I_i^+I_j^-,\hspace{2.9cm} (2.1.23)
\end{eqnarray*}
where
$$\theta_{i,j}=\delta_{i,j-1}-\delta_{i,j+1},\;\;\varepsilon_{i,j}=\left\{ \matrix{1 & i<j,\cr 0 & i=j, \cr -1
& i>j.}\right. \eqno (2.1.24)$$ Essler, Korepin and Schoutens
found the complete solutions ([29-30]) and excitation spectrum
([31]) of 1-D Hubbard model chain.

 $\bullet$ Haldane-Shastry model ([19-21]) whose Hamiltonian is
 given by a family. The first member is
$$H_2={\sum_{i,j}}'(\frac{Z_iZ_j}{Z_{ij}Z_{ji}})(P_{ij}-1),\eqno (2.1.25)$$
where and henceforth the ' stands for $i\ne j$ in the summation
and $P_{ij}=2({\bf S}_i\cdot {\bf S}_j+\frac{1}{4})$,
$Z_k=\exp^{i\pi\frac{k}{N}}$, $Z_{ij}=Z_i-Z_j$. The next reads
$$H_3={\sum_{i,j,k}}'(\frac{Z_iZ_jZ_k}{Z_{ij}Z_{jk}Z_{ki}})(P_{ijk}-1),\eqno
(2.1.26)$$ and
$$H_4={\sum_{i,j,k,l}}'(\frac{Z_iZ_jZ_kZ_l}{Z_{ij}Z_{jk}Z_{kl}Z_{li}})(P_{ijkl}-1)+H_4',\eqno
(2.1.27)$$
$$H_4'=-\frac{1}{3}H_2-2{\sum_{i,j}}'(\frac{Z_iZ_j}{Z_{ij}Z_{ji}})^2(P_{ij}-1),\eqno
(2.1.28)$$ where
\begin{eqnarray*}
 \hspace{2cm} P_{ijk}&=&P_{ij}P_{jk}+P_{jk}P_{ki}+P_{ki}P_{ij},\\
P_{ijkl}&=&P_{ij}P_{jk}P_{kl}+({\rm cyclic}\;\; {\rm
for}\;\;i,j,k\;\;{\rm and}\;\; l). \hspace{4cm} (2.1.29)
\end{eqnarray*}
The eigenvalues of $H_2$ and $H_3$ have been solved in Ref. [21]
and numerical calculations were made for $H_4$. The $H_2$ and
$H_3$ were shown to be obtained in terms of quantum determinant
([32]).

$\bullet$ Hydrogen atom (with and without monopole, [33])

$$H=\frac{{\bf \pi}^2}{2\mu}+\frac{1}{2\mu}\frac{q^2}{r^2}-\frac{\kappa}{r},\ \ \
{\bf \pi}=p-zeA \eqno(2.1.30)$$ where $\mu$ is mass, $q=zeg$,
$\kappa=ze^2$ and $g$ being monopole charge.

 $\bullet$ Super Yang-Mills Theory ($N=4$): $Y(SO(6))$ ([34])
$$H=2\sum_\alpha\sum_jh(j)P^j_{\alpha\alpha+1},\ \
h(j)=\sum^j_{k=1}\frac{1}{k},\; h(0)=1.\eqno (2.1.31)$$
 where $P^j$ is projector for the weight $j$ of $SU(2)$ and $\alpha$ stands for
``lattice'' index.

\bigskip

\subsection{$Y(SU(3))$}

For the Yangian associated to $SU(3)$, there are the following
independent relations
$$[I_\lambda,I_\mu]=if_{\lambda\mu\nu}I_\nu,\ \
[I_\lambda,J_\mu]=if_{\lambda\mu\nu}J_\nu\ \
(\lambda,\mu,\nu=1,\cdots,8).\eqno (2.2.1)$$ Define
$$I^{(1)}_\pm=I_1\pm iI_2,\ U_\pm^{(1)}=I_6\pm
iI_7,\ V^{(1)}_\pm =I_4\mp i I_5,\
\frac{\sqrt{3}}{2}I^{(1)}_8=I_8.\eqno (2.2.2)$$ and $J_\mu$
represents the corresponding operator for
$I^{(2)}_\pm,U^{(2)}_\pm,V^{(2)}_\pm$ and $I^{(2)}_8,I^{(2)}_3$.
After lengthy calculation one finds that based on RTT relation
there is only one independent relation for $Y(SU(3))$ additional
to equation (2.2.1):
$$
[I_8^{(2)},I^{(2)}_3]=\frac{1}{3!}(\{I^{(1)}_+,U^{(1)}_+,V^{(1)}_+\}-\{I_-^{(1)},U^{(1)}_-,V^{(1)}_-\})\eqno
(2.2.3)$$ where $\{\cdots\}$ stands for the symmetric summation.
The conclusion can be verified through both the Drinfel'd formula
$(C_{\lambda\mu\nu}=if_{\lambda\mu\nu})$ and RTT relations with
replacing $P_{12}$ in $SU(2)$ by
$$P_{12}=\frac{1}{3}I+\frac{1}{2}\sum_\mu\lambda_\mu\lambda_\mu,\eqno
(2.2.4)$$ where $\lambda_\mu$ are the Gell-Mann matrices. Setting
$$
T(u)=\sum^\infty_{n=0}u^{-n}T^{(n)},\eqno (2.2.5)$$
$$T^{(n)}= {\left
[\matrix{\frac{1}{3}T^{(n)}_0+T^{(n)}_3+\frac{1}{\sqrt{3}}T^{(n)}_8
& T^{(n)}_1-iT^{(n)}_2 & T^{(n)}_4-iT^{(n)}_5\cr
T^{(n)}_1+iT^{(n)}_2 &
\frac{1}{3}T^{(n)}_0-T^{(n)}_3+\frac{1}{\sqrt{3}}T^{(n)}_8 &
T^{(n)}_6-iT^{(n)}_7\cr T^{(n)}_4+iT^{(n)}_5 &
T^{(n)}_6+iT^{(n)}_7 &
\frac{1}{3}T^{(n)}_0-\frac{2}{\sqrt{3}}T^{(n)}_8 }\right]},\eqno
(2.2.6)$$ and substituting them into RTT relation we find
equations (2.2.1)-(2.2.3) are independent relations together with
the co-product, for example,
\begin{eqnarray*}
\hspace{2.5cm}\Delta I^{(2)}_\pm & = & I^{(2)}_\pm\otimes
1+1\otimes I^{(2)}_\pm
 \pm 2(I^{(1)}_3\otimes I^{(1)}_\pm-I^{(1)}_\pm\otimes
I^{(1)}_3)\\
&+&\frac{1}{2}(V^{(1)}_\mp\otimes U^{(1)}_\mp - U^{(1)}_\mp\otimes
V^{(1)}_\mp)\hspace{4.9cm} (2.2.7)
\end{eqnarray*}
and others.

The quantum determinant of $T(u)$ which is $3$ by $3$ matrix for
the fundamental representation of $gl(3)$ takes the form
\begin{eqnarray*}
\hspace{1.5cm}\tilde{\det}_3T(u)&=& T_{11}(u)\{
T_{22}(u-1)T_{33}(u-2)-T_{23}(u)T_{32}(u-2)\}\\
&\mbox{}& -T_{12}(u)\{
T_{21}(u-1)T_{33}(u-2)-T_{23}(u-1)T_{31}(u-2)\}\\
&\mbox{}& +T_{13}(u)\{
T_{21}(u-1)T_{32}(u-2)-T_{22}(u-1)T_{31}(u-2)\}\\
&=& \sum_p(-1)^p T_{1p_1}(u)T_{2p_2}(u-1)T_{3p_3}(u-2)
\hspace{3.8cm} (2.2.8)
\end{eqnarray*}
where $p$ stands for all the possible arrangements of
$(p_1,p_2,p_3)$. In comparison with the quantum determinant
$$\tilde{\det}_2T(u)=\sum_{k,l,m=0}^\infty \frac{(l-m-1)!}{(m-1)!l!}
u^{-(m+l+k)}(T_{11}^{(k)}T_{22}^{(m)}-T_{12}^{(k)}T_{21}^{(m)}),\eqno
(2.2.9)$$ now we have
\begin{eqnarray*}
\hspace{1.5cm}\tilde{\det}_3T(u)&=&\sum_{k,l,m,p,q=0}^\infty
\frac{(l+m-1)!}{(m-1)!l!}\frac{2^q(p+q-1)!}{(p-1)!q!}
u^{-(m+l+k+p+q)}\\
&\mbox{}&\{T_{11}^{(k)}(T_{22}^{(m)}T_{33}^{(p)}-T_{23}^{(m)}T_{32}^{(p)})
-T_{12}^{(k)}(T_{21}^{(m)}T_{33}^{(p)}-T_{23}^{(m)}T_{31}^{(p)})\\
&\mbox{}&+T_{13}^{(k)}(T_{21}^{(m)}T_{32}^{(p)}-T_{22}^{(m)}T_{31}^{(p)})\}\\
&=& \sum_{n=0}^\infty u^{-n}C_n,\hspace{8.2cm} (2.2.10)
\end{eqnarray*}
i.e.,
$$C_0=1, C_1=T_0^{(1)},
C_2=T_0^{(2)}+T_0^{(1)}+2(T_0^{(1)})^2-{\bf I}^2,\eqno (2.2.11)$$
$${\bf I}^2=\sum_{\lambda=1}^\infty {\bf I}_{\lambda}^2.\eqno (2.2.12)$$
When we constrain $\tilde {\det} T(u)=1$ it leads to $Y(SU(2))$
and $Y(SU(3))$ that are formed by the set $\{ I_\lambda,
J_\lambda\}$, $\lambda=1,2,3$ and $\lambda=1,2,\cdots,8$ for
$SU(2)$ and $SU(3)$, respectively.

\vskip 0.5cm

An example of realization of $Y(SU(3))$ is the generalization of
Haldane-Shastry model ([19-21]) for the fundamental representation
of generators of $SU(3)$:
$$I_\mu=\sum_iF^\mu_i,\eqno (2.2.13)$$
$$J_\mu=\sum_i\mu_iF^\mu_i+\lambda f_{\mu\lambda\nu}\sum_{i\neq
j}W_{ij}F^\nu_iF^\lambda_j,\eqno (2.2.14)$$ where $W_{ij}$
satisfies the same relation as in Haldane-Shastry model given in
section 2.1 and $F^\mu$ are the Gell-Mann matrices.

\bigskip

\subsection{$Y(SO(5))$ and $Y(SO(6))$}

 For $SO(N)$ it holds
$$[L_{ij},L_{kl}]=iC^{st}_{ij,kl}L_{st},\eqno (2.3.1)$$
where
$$C^{st}_{ij,kl}=\delta_{ik}\delta_{js}\delta_{lt}-\delta_{il}\delta_{js}\delta_{kt}-\delta_{jk}\delta_{is}\delta_{lt}
+\delta_{jl}\delta_{is}\delta_{kt}.\eqno (2.3.2)$$

The rational solutions of YBE for $SO(N)$ were firstly given by
Zamolodchikov's ([35]). They are also re-derived by taking the
rational limit of the trigonometric R-Matrix:
$$
\breve{R}(u)=f(u)[u^2P+u(A-I-\frac{3}{2}P)\xi+\frac{3}{2}I\xi^2],\eqno
(2.3.3)$$ where $u$ stands for spectral parameter and $\xi$ the
other free parameter ([36-37]). The elements of $\breve{R}(u)$ are
($a,b,c,d=-2,-1,0,1,2$)
$$
[\breve{R}(u)]^{ab}_{cd}=u^2\delta_{ab}\delta_{bc}+u(\delta_{a-b}\delta_{c-d}-\delta_{ac}\delta_{bd}-
\frac{3}{2}\delta_{ad}\delta_{bc})\xi
+\frac{3}{2}\delta_{ac}\delta_{bd}\xi^2.\eqno (2.3.4)$$

 For $SO(5)$, we introduce
$$T^{(1)}=\xi \left[ \matrix{E_3-\frac{3}{2} & U_+ & E_+ & V_+
& 0 \cr U_- & F_3-\frac{3}{2} & F_+ & 0 & -V_+ \cr E_- & F_- &
-\frac{3}{2} & -F_+ & -E_+\cr V_-  & 0 & -F_- & -F_3-\frac{3}{2} &
-U_+ \cr 0 & -V_- & -E_- & -U_- &-E_3-\frac{3}{2} }\right],\eqno
(2.3.5)$$ where $$\mbox{}\hspace{0.8cm}
\matrix{E_3=E_{22}-E_{-2,-2}, & F_3=E_{11}-E_{-1-1}, &
U_+=E_{21}-E_{-1-2}, \cr V_+=E_{2-1}-E_{1-2}, &
E_+=E_{20}-E_{0,-2}, & F_+=E_{10}-E_{0-1},\cr U_-=E_{12}-E_{-2-1},
& V_-=E_{-12}-E_{-2} & E_-=E_{02}-E_{-20}, \cr
F_-=E_{01}-E_{-10;}& & \cr }\eqno (2.3.6)$$
$$T^{(2)}_{ab}=\frac{3}{2}\xi^2 E^{(2)}_{ab}\ \ \ \
(a,b=-2,-1,0,1,2).\eqno (2.3.7)$$ Substituting $T^{(n)}$ (only
$n=1,2$ are needed to be considered)
 into RTT relation, there appears 35 relations for $J_\mu$ besides
 the Jacobi identities. However , a lengthy computation shows that
 besides
$$\matrix{[I_\alpha,I_\beta]=C^\gamma_{\alpha\beta}I_\gamma\cr
[I_\alpha,J_\beta]=C^\gamma_{\alpha\beta}J_\gamma } \ \ \ \
(\alpha=i,j),\eqno (2.3.8)$$
 there is only one independent relation
$$
[E^{(2)}_3,F^{(2)}_3]=\frac{1}{4!}(\{U_-,E_+,F_-\}-\{U_+,E_-,F_+\}-\{V_+,E_-,F_-\}+\{V_-,E_+,F_+\}),\eqno
(2.3.9)$$where again $\{\ \  \}$ stands for the symmetric
summation.

\vskip 0.5cm

A realization of $Y(SO(5))$is given as follows. Set
$$I_{ab}(x)=\frac{1}{2}\psi^+_\alpha(x)(I^{ab})_{\alpha\beta}\psi_\beta(x)\
\ \ \ \ (a,b=-2,-1,0,1,2),\eqno (2.3.10)$$
$$\{\psi^+_\alpha(x),\psi_\beta(y)\}_+=\delta(x-y)\delta_{\alpha\beta}.\eqno
(2.3.11)$$ Then $$ I_{ab}=\sum\limits_x I_{ab}(x),\eqno (2.3.12)$$
$$J_{ab}=\sum\limits_{x,y,c\ne a,b}\epsilon(x-y) I_{ac}(x)
I_{cb}(y)\eqno (2.3.13)$$ satisfies the commuting relations for
$Y(SO(5))$. The following Hamiltonian of ladder model  not only
commutes with $I_{ab}$, i.e., it possesses $SO(5)$ symmetry, but
also commutes with $J_{ab}$.
\begin{eqnarray*}
\hspace{2cm}H &=& H_1 + \sum\limits_x H_2(x) + \sum\limits_x H_3(x);\hspace{6.1cm} (2.3.14)\\
H_1 &=& 2 t_1 \sum\limits_{<x,y>} [c_{\sigma}^{+} (x) c_{\sigma}
(y)+d_{\sigma}^{+} (x) d_{\sigma} (y) + H.C.];\hspace{3.6cm} (2.3.15)\\
H_2(x) &=& U (n_{c\uparrow} - \frac{1}{2})(n_{c\downarrow} -
\frac{1}{2}) + (c\rightarrow d)+ V (n_c - 1)(n_d - 1) + J {\bf
S}_c \cdot {\bf S}_d\\
&=&\frac{J}{4}\sum\limits_{a<b} I_{ab}^2 + ( \frac{1}{8} J +
\frac{1}{2} U) (\psi_{\alpha}^+ \psi_{\alpha} - 2);\hspace{4.9cm} (2.3.16)\\
H_3 (x) &=& - 2 t_3 (c_{\sigma}^+ (x)d_{\sigma} (x) +
H.C.).\hspace{6.5cm} (2.3.17)
\end{eqnarray*}

\vskip 0.5cm

Because locally $SO(6)\simeq SU(4)$ we introduce ($15$ generators)
$$T_{ab}^{(1)} = I_{ab},\;\;T_{ab}^{(2)} = I_{ab}^{(2)}\\
(a,b=1,2,\ldots,6.).\eqno (2.3.18)$$ and the $\check{R}(u)$-matrix
reads
$$\check{R}(u) = f(u) [ u^2 P + u \xi ( A - 2P - I ) + 2 \xi^2
I].\eqno (2.3.19)$$

The RTT relation gives $4+4+441+315+225$ more relations. After
careful calculations one finds ([15-16]) that there are the
following independent relations for $J_{ab}$ themselves:
\begin{eqnarray*}
\hspace{2cm}{[I_{12}^{(2)},I_{34}^{(2)}}]&=&\frac{i}{24}(\{I_{23},I_{16},I_{46}\}+\{I_{23},
I_{15},I_{45}\}+\{I_{14},I_{25},I_{35}\}\\
&\ &+\{I_{14},I_{26},I_{36}\}-\{I_{13},I_{26},I_{46}\}-\{I_{13},I_{25},I_{45}\}\\
&\ &-\{I_{24},I_{15},I_{35}\}-\{I_{24},I_{16},I_{36}\});\hspace{4cm} (2.3.20)\\
{[I_{12}^{(2)},I_{56}^{(2)}}]&=&\frac{i}{24}(\{I_{15},I_{23},I_{36}\}+\{I_{15},I_{24},
I_{46}\}+\{I_{26},I_{13},I_{35}\}\\
&\ &+\{I_{26},I_{14},I_{45}\}-\{I_{25},I_{13},I_{36}\}-\{I_{25},I_{14},I_{46}\}\\
&\ &-\{I_{16},I_{23},I_{35}\}-\{I_{16},I_{24},I_{45}\});\hspace{4cm} (2.3.21)\\
{[I_{34}^{(2)},I_{56}^{(2)}}]&=&\frac{i}{24}(\{I_{45}^{(1)},I_{13}^{(1)},
I_{16}^{(1)}\}+\{I_{45}^{(1)},I_{23}^{(1)},I_{26}^{(1)}\}+\{I_{36}^{(1)},
I_{14}^{(1)},I_{16}^{(1)}\}\\
&\ &+\{I_{36}^{(1)},I_{24}^{(1)},I_{26}^{(1)}\}
-\{I_{35}^{(1)},I_{14}^{(1)},I_{16}^{(1)}\}-\{I_{35}^{(1)},I_{24}^{(1)},
I_{26}^{(1)}\}\\
&\ &-\{I_{46}^{(1)},I_{13}^{(1)},I_{16}^{(1)}\}-\{I_{46}^{(1)},
I_{23}^{(1)},I_{26}^{(1)}\}).\hspace{3.4cm} (2.3.22)
\end{eqnarray*}

\bigskip

\section{Applications of Yangian}

The first example was given by Belavin ([38]) in deriving the
spectrum of nonlinear $\sigma$ model. Here we only show briefly
some interpretations of Yangian through the particular
realizations of Yangian.
\bigskip
\subsection{Reduction of $Y(SU(2))$}
The simplest realization of $Y(SU(2))$ is made of two-spin system
with $\bf{S_1}$ and $\bf{S_2}$ (any dimensional representations of
$SU(2)$):
$$\bf{J'}=\frac{1}{\mu+\nu}\bf{J}=\frac{1}{\mu+\nu}(\mu\bf{S_1}\times {\bf 1}+\nu\bf{S_2}\times {\bf 1}+2\lambda\bf{S_1}\times\bf{S_2}),\eqno
(3.1.1)$$ that contains the (antisymmetric) tensor interaction
between $\bf{S_1}$ and $\bf{S_2}$. For example, for Hydrogen atom
$\bf{S_1}=\bf{L}$ and $\bf{S_2}=\bf{K}$ (Lung-Lenz vector).

For $S_1=S_2=1/2$, when $$\mu\nu=\lambda^2,\eqno (3.1.2)$$ we
prove that after the following similar transformation
$${\bf Y}=A{\bf J'}A^{-1},\;\;
A=\left[\begin{array}{llcl}1&0&0&0\\0&{\nu}&{i\lambda}&0\\0&{i\lambda}&{\nu}&0\\0&0&0&1\end{array}\right],\eqno
(3.1.3)$$ the Yangian reduces to $SO(4)$:\
($\rho=\nu+i\lambda=\sqrt{\nu^2+\lambda^2}e^{i\theta}$)
\begin{eqnarray*}
\hspace{2.7cm}&&Y_1=\left[\begin{array}{llcl}M_1&0\\0&L_1\end{array}\right],\ M_1=\frac{1}{2}\left[\begin{array}{llcl}0&{\rho}\\{\rho^{-1}}&0\end{array}\right],\ L_1=\frac{1}{2}\left[\begin{array}{llcl}0&{\rho^{-1}}\\{\rho}&0\end{array}\right],\\
&&Y_2=\left[\begin{array}{llcl}M_2&0\\0&L_2\end{array}\right],\ M_2=\frac{1}{2}\left[\begin{array}{llcl}0&{-i\rho}\\{i\rho^{-1}}&0\end{array}\right],\ L_2=\frac{1}{2}\left[\begin{array}{llcl}0&{-i\rho^{-1}}\\{i\rho}&0\end{array}\right],\\
&&Y_3=\left[\begin{array}{llcl}{\frac{1}{2}}{\sigma_3}&0\\0&{\frac{1}{2}}{\sigma_3}\end{array}\right],\
M_3=\frac{1}{2}\sigma_3. \hspace{5.5cm}(3.1.4)
\end{eqnarray*}
and
$${\bf Y}^2=\frac{1}{2}(\frac{1}{2}+1)=\frac{3}{4}.\eqno (3.1.5)$$
Namely, under $\mu\nu=\lambda^2$, the ${\bf Y}$ reduces to $SO(4)$
by $M_\pm=M_1\pm{iM_2}$, $M_+=\rho\sigma_+$,
$M_-=\rho^{-1}\sigma_-$. The scaled $M_\pm$ and $M_3$ still
satisfy the $SU(2)$ relations: $$[M_3,M_\pm]=\pm{M_\pm},\;\;
{[M_+,M_-]}=2M_3.\eqno (3.1.6)$$ and there are the  similar
relations for ${\bf L}$.

It should be emphasized that here the new ``spin" ${\bf M}$ (and
${\bf L}$) is the consequence of two spin($\frac{1}{2}$)
interaction. As usual for two 2-dimensional representations of
$SU(2)$ (Lie algebra)
$$\underline{2}\otimes\underline{2}=\underline{3}\;\;({\rm spin\
triplet})\oplus\underline{1}\;\;({\rm singlet}).\eqno (3.1.7)$$
However, here we meet a different decomposition:
$$\underline{2}\otimes\underline{2}=\underline{2}({\bf
M})\oplus\underline{2}({\bf L}).\eqno (3.1.8)$$

\vskip 0.5cm

The idea can be generalized to $SU(3)$'s fundamental
representation
$$J_\lambda=uI_1^\lambda+vI_2^\lambda+\lambda{f_{\lambda\mu\nu}}
\sum\limits_{i<j}F_{1i}^\mu{F_{2j}^\nu},\eqno (3.1.9)$$
$${[F_{i\mu},F_{j\nu}]}=if_{\mu\nu\lambda}F_{i\lambda}\delta_{ij}\ \ \ \ \
(\lambda,\mu,\nu=1,2,\cdots,8).\eqno (3.1.10)$$ Under the
condition
$$uv=\lambda^2,\ \ \ \ \ \,\ \ \ v+i\lambda=\rho,\eqno (3.1.11)$$ and
the similar transformation
$$ Y_\mu=AJ_\mu{A^{-1}/(u+v)},\;\;
A=\left[\begin{array}{ccccccccc}1&0&0&0&0&0&0&0&0\\0&{\nu}&0
&i{\lambda}&0&0&0&0&0\\0&0&{\nu}&0&0&0&i{\lambda}&0&0\\0&i{\lambda}
&0&{\nu}&0&0&0&0&0\\0&0&0&0&1&0&0&0&0\\0&0&0&0&0&{\nu}&0&i{\lambda}
&0\\0&0&i{\lambda}&0&0&0&{\nu}&0&0\\0&0&0&0&0&i{\lambda}&0&{\nu}
&0\\0&0&0&0&0&0&0&0&1\end{array}\right],\eqno (3.1.12)$$ the
Yangian then reduces to
\begin{eqnarray*}
\hspace{1cm}&&Y(I_-)=\left[\begin{array}{ccc}{\rho^{-1}}I_-&0&0\\0&{\rho}I_-&0\\0&0&I_-\end{array}\right],\
Y(I_+)=\left[\begin{array}{ccc}{\rho}I_+&0&0\\0&{\rho^{-1}}I_-&0\\0&0&I_3\end{array}\right],\\
&&Y(I_8)=\frac{\sqrt3}{3}\left[\begin{array}{ccc}{\lambda_3}&0&0\\0&{\lambda_3}&0\\0&0&{\lambda_3}\end{array}\right],\
Y(I_3)=\frac{1}{2}\left[\begin{array}{ccc}{\lambda_3}&0&0\\0&{\lambda_3}&0\\0&0&{\lambda_3}\end{array}\right],\\
&&Y(U_+)=\left[\begin{array}{ccc}{U_+}&0&0\\0&{\rho}U_+&0\\0&0&{\rho^{-1}}U_+\end{array}\right],\
Y(U_-)=\left[\begin{array}{ccc}{U_-}&0&0\\0&{\rho^{-1}}U_-&0\\0&0&{\rho}U_-\end{array}\right],\\
&&Y(V_+)=\left[\begin{array}{ccc}{\rho^{-1}}{V_-}&0&0\\0&V_-&0\\0&0&{\rho}V_-\end{array}\right],\
Y(V_-)=\left[\begin{array}{ccc}{\rho}{V_-}&0&0\\0&V_-&0\\0&0&{\rho^{-1}}V_-\end{array}\right].\hspace{1.3cm} (3.1.13)\\
\end{eqnarray*}
The usual decomposition through the Clebsch-Gordan coefficients
for the representations of Lie algebra $SU(3)$ is
$\underline{3}\otimes\underline{3}=\underline{6}\oplus\underline{3}$.
However, here we have $$
\underline{3}\otimes\underline{3}=\underline{3}\oplus\underline{3}\oplus\underline{3},\eqno
(3.1.14)$$ and $$
\sum\limits_{\lambda=1}^8Y_\lambda^2=\frac{1}{u+v}\sum\limits_{\lambda=1}^\infty{J_\lambda^2}=\frac{1}{3}.\eqno
(3.1.15)$$ It is easy to check that the rescaling factor $\rho$
does not change the commutation relations for $SU(3)$ formed by
$I_\pm$, $U_\pm$, $V_\pm$, $I_3$ and $I_8$. In general, we guess
for the fundamental representation of $SU(n)$ we shall meet
$$\underline{n}\otimes
\underline{n}=\underline{n}\oplus\underline{n}\oplus\underline{n}+\cdots+\underline{n}\
\ (n\;\; {\rm times}).\eqno (3.1.16)$$

\vskip 0.5cm

Next we consider Yang-Mills gauge field for reduced $Y(SU(2))$.
For a tensor wave function $(x\equiv
\{x_{1},x_{2},x_{3},x_{0}\})$,
$$\Psi(x)=\|\psi_{ij}(x)\|\ \ (i,j=1,2,3,4).\eqno (3.1.17)$$ An
isospin transformation yields
$$\Psi'(x)=U(x)\Psi(x),\;\;U(x)=1-i\theta^{a}J_{a},\eqno (3.1.18)$$
where $$J^{a}=uS_{a}\otimes\textbf{1}+v\textbf{1}\otimes
S_{a}+2\lambda\epsilon_{abc} S^{b}\otimes S^{c},\eqno (3.1.19)$$
or
$$[J_{a}]^{\alpha\beta}_{\gamma\delta}=u(S^{a})_{\alpha\gamma}\delta_{\beta\delta}
+v(S^{a})_{\beta\delta}\delta_{\alpha\gamma}+i\alpha\varepsilon_{abc}
(S^{b})_{\alpha\gamma}(S^{c})_{\beta\delta}.\eqno (3.1.20)$$
Define
$$
D_{\mu}=\partial_{\mu}+gA_{\mu},\eqno (3.1.21)$$ i.e.,
$$[D_{\mu}\psi]_{\alpha\beta}=\partial_{\mu}\psi_{\alpha\beta}+
gA^{a}_{\mu}[Y_{a}]^{\alpha\beta}_{\gamma\delta}\psi_{\gamma\delta}(x),\;\;A_{\mu}=A^{a}_{\mu}J_{a}.\eqno
(3.1.22)$$ The gauge-covariant derivative should preserve
$$\delta(D_{\mu}\psi)=0,\eqno (3.1.23)$$
i.e., $$(-i\partial_{\mu}\theta^{a}(x)+g\delta A^{a}_{\mu})
[Y_{a}]^{\alpha\beta}_{\gamma\delta}-ig\theta^{a}(x)A^{b}_{\mu}
[J_{b},J_{a}]^{\alpha\beta}_{\gamma\delta}=0.\eqno (3.1.24)$$ When
$uv=\lambda^{2}$ and by rescaling
$$Y_{a}=(u+v)J_{a},\eqno (3.1.25)$$
we have $$\delta
A^{a}_{\mu}=\epsilon_{abc}\theta^{b}(x)A^{c}_{\mu}(x)+\frac{i}{g}\partial_{\mu}
\theta^{a}(x),\eqno (3.1.26)$$ and $$
F_{\mu\nu}=\frac{1}{g}[D_{\mu},D_{\nu}]=F^{a}_{\mu\nu}Y_{a},\eqno
(3.1.27)$$
$$F^{a}_{\mu\nu}=\partial_{\mu}A^{a}_{\gamma}-\partial_{\nu}A^{a}_{\mu}
+ig\epsilon_{abc}A^{b}_{\mu}A^{c}_{\gamma}.\eqno (3.1.28)$$ Here
the tensor isospace has been separated to two irrelevant spaces,
i.e., $\Psi={\left[\matrix{\Psi_1 & 0 \cr 0 & \Psi_2} \right]}$
where $\Psi_1$ and $\Psi_2$ are $2\times2$ wavefunction.

\bigskip

\subsection{Illustrative examples: NMR of Breit-Rabi Hamiltonian
and Yangian}

The Breit-Rabi Hamiltonian is given by
$$H={\bf K}\cdot {\bf S}+\mu{\bf B}\cdot
{\bf S},\eqno (3.2.1)$$ where  $S=\frac{1}{2}$ and $B={\bf B}(t)$
is magnetic field.

The Hamiltonian can easily be diagonalized for any background
angular momentum (or spin) ${\bf K}$. The $ {\bf S}$ stands for
spin of electron and for simplicity ${\bf K}={\bf
S_{1}}$($S_{1}=1/2$) is an average background spin contributed by
other source, say, control spin. Denoting by
$$H=H_{0}+H_{1}(t),\;\; H_{0}=\alpha{\bf S_{1}}\cdot {\bf
S_{2}},\;\;H_{1}(t)=\mu{\bf B}(t)\cdot{\bf S_{2}}.\eqno (3.2.2)$$

 Let us work in the interaction picture:
$$H_{I}=\mu{\bf B}(t)\cdot (e^{i\alpha{\bf
S_{1}}\cdot{\bf S_{2}}}{\bf S_{2}} e^{-i\alpha{\bf S_{1}}\cdot{\bf
S_{2}}}) = \mu{\bf B}(t)\cdot{\bf J},\eqno (3.2.3)$$
$${\bf
J}=\mu_{1}{\bf S_{1}}+\mu_{2}{\bf S_{2}} +2\lambda({\bf
S_{1}}\times{\bf S_{2}}),\eqno (3.2.4)$$ where
$\mu_{1}=\frac{1}{2}(1-cos\alpha)$,
$\mu_{2}=\frac{1}{2}(1+cos\alpha)$,
$\lambda=\frac{1}{2}sin\alpha$. Obviously, here we have
$\mu_{1}\mu_{2}=\lambda^{2}$. It is not surprising that the
$Y(SU(2))$ reduces to $SO(4)$ here because the transformation is
fully Lie-algebraic operation. This is an exercise in quantum
mechanics.

\vskip 0.5cm

 For generalization we regard $\mu_{1}$ and $\mu_{2}$
as independent parameters,i.e., drop the relation
$\mu_{1}\mu_{2}=\lambda^{2}$. Looking at $${\bf J}=\mu_{1}{\bf
S_{1}}+\mu_{2}{\bf S_{2}} -\frac{1}{2}(\mu_{1}+\mu_{2})({\bf
S_{1}}+{\bf S_{2}})+ \gamma({\bf S_{1}}+{\bf S_{2}})+2\lambda {\bf
S_{1}}\times{\bf S_{2}}.\eqno (3.2.5)$$

When $\gamma=\frac{1}{2}$, $\mu_{2}-\mu_{1}=cos\alpha$ and
$\lambda=\frac{1}{2}sin\alpha$, it reduces to the form in the
interacting picture. Putting $${\bf S_{1}}+{\bf S_{2}}={\bf
S},\;\;2\lambda=-\frac{h}{2} (h\;\;{\rm is}\;\;{\rm not}\;\;{\rm
Plank}\;\;{\rm constant)}.\eqno (3.2.6)$$
 In accordance with the convention we have
$${\bf J}=\gamma{\bf S}+\sum^{2}_{i=1}\mu_{i} {\bf
S_{i}}+\frac{h}{2}{\bf S_{1}}\times{\bf S_{2}}
-\frac{1}{2}(\mu_{1}+\mu_{2}){\bf S}=\gamma{\bf S} +{\bf Y}.\eqno
(3.2.7)$$ Since ${\bf J}\rightarrow\xi{\bf S}+{\bf J}$ still
satisfies Yangian relations, it is natural to appear the term
$\gamma{\bf S}$. The interacting Hamiltonian then reads
$$H_{I}(t)=-\gamma{\bf B}(t)\cdot{\bf S}- {\bf B}(t)\cdot{\bf
Y}.\eqno (3.2.8)$$

When $\mu_{i}=0$, $h=0$, it is the usual NMR for spin 1/2. To
solve the equation, we use $$i\frac{\partial\Psi(t)}{\partial
t}=H_{I}(t)\Psi(t),\;\;|\Psi(t)\rangle=\sum_{\alpha=\pm,3;0}a_{\alpha}(t)|\chi_{\alpha}\rangle,\eqno
(3.2.9)$$ where $\{\chi_{\pm},\chi_{3}\}$ is the spin triplet and
$\chi_{0}$ singlet. Setting
$$B_{\pm}(t)=B_1{(t)}\pm iB_{2}(t)=B_{1}e^{\mp i\omega_{0}t},\;\;{\rm and}\;\;B_{3}={\rm const}.\eqno (3.2.10)$$
and rescaling by $$a_{\pm}(t)=e^{\pm i\omega_{0}t}b_{\pm(}t),\eqno
(3.2.11)$$ we get
\begin{eqnarray*}\hspace{1.8cm}
i\frac{db_{\pm}(t)}{dt}&=&-\gamma\{\frac{1}{\sqrt{2}}B_{1}a_{3}(t)
\mp(\omega_{0}\gamma^{-1}-B_{3})b_{\pm}(t)\}\pm\frac{1}{2\sqrt{2}}\mu_{-}B_{1}a_{0}(t),\\
i\frac{da_{3}(t)}{dt}&=&-\frac{\gamma
B_{1}}{\sqrt{2}}\{b_{+}(t)+b_{-}(t)\}-\frac{1}{2}
\mu_{-}B_{3}a_{0}(t),\\
i\frac{da_{0}(t)}{dt}&=&-\frac{1}{2}\mu_{+}\{\frac{1}{\sqrt{2}}
B_{1}[b_{-}(t)-b_{+}(t)]\}+B_{3}a_{3}(t),\hspace{3.2cm} (3.2.12)
\end{eqnarray*}
where $\mu_{\pm}=(\mu_{1}-\mu_{2}\pm i\frac{h}{2})$, i.e.,
$$|\Phi(t)\rangle=\left[\begin{array}{c}b_{1}(t)\\
a_{3}(t)\\
b_{-}(t)\\
a_{0}(t)
\end{array}\right],\mathcal{H}_{I}=\left[\begin{array}{cccc}
\omega_{0}-\gamma B_{3}&-\gamma B_1\frac{1}{\sqrt{2}}&0&\frac{1}{2\sqrt{2}}\mu_{-}B_{1}\\
-\gamma B_1\frac{1}{\sqrt{2}}&0&-\gamma B_1\frac{1}{\sqrt{2}}&-\frac{1}{2}\mu_{-}B_{3}\\
0&-\gamma B_1\frac{1}{\underline{\sqrt{2}}}&-(\omega_{0}-\gamma B_{3})&-\frac{1}{2\sqrt{2}}\mu_{-}B_{1}\\
\frac{1}{2\sqrt{2}}\mu_{+}B_{1}&-\frac{1}{2}\mu_{+}B_{3}&-\frac{1}{2\sqrt{2}}\mu_{+}B_{1}&0
\end{array}\right],\eqno (3.2.13)$$
$$i\frac{d|\Phi(t)\rangle}{dt}={H_{I}}|\Phi(t)\rangle.\eqno
(3.2.14)$$ Noting that $\mathcal{H_{I}}$ is independent of time,
we get $$|\Phi(t)\rangle=e^{-iEt}|\Phi(t)\rangle.\eqno (3.2.15)$$
Then
$$\det|{H_{I}}-E|=0\eqno (3.2.16)$$
leads to $$E^{4}-[(\omega_{1}-\gamma
B_{3})^{2}+\gamma^{2}B_{1}^{2}+\frac{1}{4}
\mu_{+}\mu_{-}(B^{2}_{1}+B^{2}_{3})]E^{2}+$$
$$\frac{1}{4}\mu_{+}\mu_{-}[B^{2}_{3}(\omega_{0}-\gamma
B_{3})^{2}-2\gamma B_{3}B^{2}_{1}(\omega_{0}-\gamma
B_{3})+\gamma^{2}B^{4}_{1}]=0.\eqno (3.2.17)$$

There is a transition between the spin singlet and triplet in the
NMR process, i.e., the Yangian transfers the quantum information
through the evolution. The simplest case is $B_1=0$, then the
eigenvalues are
$$ E=\pm(\omega_{0}-\gamma B_{3}),
E=\pm\omega=\pm\frac{B_{3}}{2}\sqrt{(\mu_{1}-\mu_{2})^{2}+\frac{h^{2}}{4}}.\eqno
(3.2.18)$$ It turns out that there is a vibration between $s=0$
and $s=1$.
$$<s^{2}>=0\;\;{\rm at}\;\;t=\frac{\pi}{2\omega}\;\;({\rm total}\;\;{\rm spin}\;\;=0),\eqno (3.2.19)$$
$$ <s^{2}>=2\;\;{\rm at}\;\;t=\frac{\pi}{\omega}\;\;({\rm total}\;\;{\rm spin}\;\;=1).\eqno
(3.2.20)$$ Under adiabatic approximation it can be proved that it
appears Berry's phase. Obviously, only spin vector can make the
stereo angle. The role of spin singlet here is a witness that
shares energy of spin=1 state.

Actually, if
$$ B_\pm(t)=B_0\sin\theta e^{\mp i\omega_0t},\;\;B_3=B_0\cos
\theta,\eqno (3.2.21)$$ and
\begin{eqnarray*}
\hspace{2cm}&&|\chi_{11}\rangle=|\uparrow\uparrow\rangle,\;\;
|\chi_{1-1}\rangle=|\downarrow\downarrow\rangle,\;\;
|\chi_{10}\rangle=\frac{1}{\sqrt2}(|\uparrow\downarrow\rangle
+|\downarrow\uparrow\rangle),\\
&&|\chi_{00}\rangle=\frac{1}{\sqrt2}(|\uparrow\downarrow\rangle
-|\downarrow\uparrow\rangle),\hspace{7cm} (3.2.22)
\end{eqnarray*}
then let us consider the eigenvalues of
$$H=\alpha {\bf S}_1\cdot {\bf S}_2-\gamma B_0S_3-g B_0J_3,\eqno
(3.2.23)$$ under adiabatic approximation which are
$$E_\pm=\frac{1}{2}(-\frac{\alpha}{2}\pm
\sqrt{\alpha^2+g^2B_0^2\mu_+\mu_-}),\eqno (3.2.24)$$ and
$$f_1^{(\pm)}=[2(\alpha^2+g^2B_0^2\mu_+\mu _-)]^{-1/2}
[(\alpha^2+g^2B_0^2\mu_+\mu _-)^{1/2}\pm \alpha]^{1/2},\eqno
(3.2.25)$$
$$f_2^{(\pm)}=[2(\alpha^2+g^2B_0^2\mu_+\mu _-)]^{-1/2}
[\frac{\mu_+}{\mu_-}(\alpha^2+g^2B_0^2\mu_+\mu _-)^{1/2}\mp
\alpha]^{1/2}.\eqno (3.2.26)$$ We obtain the eigenstates of $H$
besides $|\chi_{1i}\rangle$ ($i=1,2$)
$$|\chi_\pm\rangle=f_1^{(\pm)}|\chi_{10}\rangle+f_2^{(\pm)}
|\chi_{00}\rangle,\eqno (3.2.27)$$ where
\begin{eqnarray*}
\hspace
{1cm}|\chi_{11}(t)\rangle&=&\cos^2\frac{\theta}{2}|\chi_{11}\rangle
+\frac{1}{\sqrt2}\sin\theta e^{-i\omega_0t}|\chi_{10}\rangle
+\sin^2\frac{\theta}{2}e^{-2i\omega_0t}|\chi_{1-1}\rangle,\\
|\chi_{1-1}(t)\rangle&=&\sin^2\frac{\theta}{2}e^{2i\omega_0t}
|\chi_{11}\rangle -\frac{1}{\sqrt2}\sin\theta
e^{i\omega_0t}|\chi_{10}\rangle
+\cos^2\frac{\theta}{2}|\chi_{1-1}\rangle,\\
|\chi_{\pm}(t)\rangle&=&\frac{1}{\sqrt2} f_1^{(\pm)} \{
-\sin\theta e^{i\omega_0t}|\chi_{11}\rangle +\sqrt
2\cos\theta|\chi_{10}\rangle +\sin\theta
e^{-i\omega_0t}|\chi_{1-1}\rangle\}\\
&\mbox{}&+f_2^{(\pm)}|\chi_{00}\rangle. \hspace {9cm} (3.2.28)
\end{eqnarray*}
We then obtain
\begin{eqnarray*}
\hspace{3cm}&&\langle\chi_{11}(t)|\frac{\partial}{\partial
t}|\chi_{11}(t)\rangle=-i\omega_0(1-\cos\theta),\\
&&\langle\chi_{1-1}(t)|\frac{\partial}{\partial
t}|\chi_{11}(t)\rangle=i\omega_0(1-\cos\theta),\\
&&\langle\chi_{\pm}(t)|\frac{\partial}{\partial
t}|\chi_{\pm}(t)\rangle=0.\hspace{6.8cm} (3.2.29)
\end{eqnarray*}
The Berry's phase is then
$$\gamma_{1\pm 1}=\pm \Omega,\;\;\Omega=2\pi (1-\cos\theta),\eqno
(3.2.30)$$ whereas $\gamma_{10}=\gamma_{00}=0$. The Yangian
changes the eigenstates of $H$, but preserves the Berry's phase.

\bigskip

\subsection{Transition between S-wave and P-wave
superconductivity}

We set for a pair of electrons:
$$S:\;\;\;\;{\rm spin}\;\;{\rm singlet},\;\;\;L=0;\eqno(3.3.1)$$
$$P:\;\;\;\;{\rm spin}\;\;{\rm triplet},\;\;\;L=1.\eqno(3.3.2)$$
Due to Balian-Werthamer ([39]), we have $$\triangle
(\textbf{k})=-\frac{1}{2}\sum_{\textbf{k}^\prime}V(\textbf{k},\textbf{k}^\prime)\frac{\triangle
(\textbf{k}^\prime)}{E(\textbf{k}^\prime)}\tanh\frac{\beta}{2}E(\textbf{k}^\prime),\eqno
(3.3.3)$$ $$E(\textbf{k})=(\epsilon^{2}(k)+|\triangle
(\textbf{k})|^{2})^{\frac{1}{2}}.\eqno (3.3.4)$$ Therefore, still
by Balian-Werthamer ([39]), we know
$$\triangle
(\textbf{k})=\triangle(k)(\frac{4\pi}{3})^{\frac{1}{2}}\left[ \begin{array}{lc} \sqrt{2}Y_{1,1}(\hat{\textbf{k}})
& Y_{1,0}(\hat{\textbf{k}})\\
 Y_{1,0}(\hat{\textbf{k}})&\sqrt{2}Y_{1,-1}(\hat{\textbf{k}}) \end{array} \right]^{*}
 =(-\sqrt{6})\triangle(k)(\frac{4\pi}{3})^{\frac{1}{2}}\Phi_{0,0}(\hat{\textbf{k}}),\eqno
 (3.3.5)$$
$$\Phi_{0,0}(\hat{\textbf{k}})
=\frac{1}{\sqrt{3}}\{Y_{1,-1}(\hat{\textbf{k}})\chi_{11}-
Y_{1,0}(\hat{\textbf{k}})\chi_{10}+Y_{1,1}(\hat{\textbf{k}})\chi_{1-1}\}
=\frac{1}{\sqrt{8}}\left[ \begin{array}{lc}\hat{\textbf{k}}_{-}
&-\hat{\textbf{k}}_{z}
 \\-\hat{\textbf{k}}_{z}&-\hat{\textbf{k}}_{+}\end{array}
 \right],\eqno (3.3.6)$$
where $\chi_{11}$,$\chi_{10}$ and $\chi_{1-1}$ stand for spin
triplet:$$\Phi_{0,0}\equiv\Phi_{J=0,m=0}.\eqno (3.3.7)$$
 The wave function of SC is
$$\phi_{0,0}=\frac{1}{\sqrt{2}}\left[
\begin{array}{lc}0 &Y_{0,0}
 \\-Y_{0,0}&0\end{array} \right].\eqno (3.3.8)$$
Introducing
$$I_{\mu}=\sum_{i=1}^{2}S_{\mu}(i);\ \ \ \ \ (\mu=1,2,3), \eqno (3.3.9)$$
$$J_{\mu}=\sum_{i=1}^{2}\lambda_{i}S_{\mu}(i)-\frac{ihv}{4}\epsilon_{\mu\lambda\nu}(S^{\lambda}(1)S^{\nu}(2)-S^{\lambda}(2)S^{\nu}(1)),\eqno
(3.3.10)$$ and noting that $J_{\mu}\rightarrow J_{\mu}+fI_{\mu}$
does not change the Yangian relations, we choose for simplicity
$f=-\frac{1}{2}(\lambda_{1}+\lambda_{2})$. Then we obtain for
$G=\hat{\textbf{k}}\cdot ({\bf J}+f{\bf I})$
$$G\phi_{0,0}=\hat{\textbf{k}}\cdot(\textbf{J}+f\textbf{I})\phi_{0,0}=\frac{\sqrt{3}}{2}
(\lambda_{2}-\lambda_{1}+\frac{hv}{2})\Phi_{0,0},\eqno (3.3.11)$$
$$G\Phi_{0,0}=\hat{\textbf{k}}\cdot(\textbf{J}+f\textbf{I})\Phi_{0,0}=\frac{1}{2\sqrt{3}}(\lambda_{2}-\lambda_{1}-\frac{hv}{2})\phi_{0,0}.\eqno
(3.3.12)$$ The transition directionally depends on the parameters
in $Y(SU(2))$. For instance,
$$SC\rightarrow PC:\ G\phi_{0,0}=\frac{\sqrt{3}}{2}\Phi_{0,0},\;
G\Phi_{0,0}=0,\;\;{\rm if}\;\;
\lambda_{1}-\lambda_{2}=-\frac{hv}{2},\eqno (3.3.13)$$ and
$$PC\rightarrow SC: \ G\phi_{0,0}=0,\;
G\Phi_{0,0}=-\frac{hv}{2\sqrt{3}}\phi_{0,0},\;\;{\rm
if}\;\;\lambda_{1}-\lambda_{2}=\frac{hv}{2}.\eqno (3.3.14)$$ We
call the type of the transition ``directional transition" ([40]).
The controlled parameters are in the Yangian operation. They
represent the interaction coming from other controlled spin.

We have got used to apply electromagnetic field $A_{\mu}$ to make
transitions between $l$ and $l\pm1$ states. Now there is Yangian
formed by two spins that plays the role changing angular momentum
states.

\bigskip

\subsection{ $Y(SU(3))$-directional transitions}

Setting
$$F_{\mu}=\frac{1}{2}\lambda_{\mu}, \
[F_{\lambda},F_{\mu}]=if_{\lambda\mu\nu}F_{\nu},\eqno (3.4.1)$$
$$I_{\mu}=\sum_{i}F_{i}^{\nu},\eqno(3.4.2)$$
$$J_{\mu}=\sum_{i}\mu_{i}F_{i}^{\mu}-ihf_{\mu\nu\lambda}\sum_{i\neq
j} W_{ij}F_{i}^{\nu}F_{j}^{\lambda},\;\;(W_{ij}=-W_{ji}),\eqno
(3.4.3)$$
$$[F_{i}^{\lambda},F_{j}^{\mu}]=if_{\lambda\mu\nu}\delta_{ij}F_i^{\nu},\eqno
(3.4.4)$$ where $\{F_{\mu}\}$ is the fundamental representation of
$SU(3)$ and $(i,j,k=1,2,...,8)$
$$\triangle_{ijk}=W_{ij}W_{jk}+W_{jk}W_{ki}+W_{ki}W_{ij}=-1.\eqno
(3.4.5)$$ (Here, no summation over repeated indices, $i\neq j\neq
k$). The reason that such a condition works only for 3-dimensional
representation of $SU(3)$ is similar to Haldane's (long-ranged)
realization of $Y(SU(2))$([19]). In $SU(2)$ long-ranged form, the
property of Pauli matrices leads to $(\sigma^{\pm})^{2}=0$.
Instead, for $SU(3)$ the conditions of $J_{\mu}$ satisfying
$Y(SU(3))$ read
$$\sum_{i\neq
j}(1-w_{ij}^2)(I_j^{+}V_i^{+}U_i^{+}-U_i^{-}V_i^{-}I_j^{-}+I_i^{+}V_j^{+}U_i^{+}
-U_i^{-}V_j^{-}I_i^-+I_j^+V_j^+U_i^+-U_i^-V_j^-I_j^-)=0,
\eqno(3.4.6)$$ and
$$\sum_i(I_i^+V_i^+U_i^+-U_i^-V_i^-I_i^-)=0,\eqno (3.4.7)$$
that are satisfied for Gell-Mann matrices.

\vskip 0.5cm

The simplest realization of $Y(SU(3))$ is then $$ W_{ij}=\left\{
\begin{array} {cl} 1\ &\ \ \ i>j\\0\ &\ \ \ i=j\\-1\ &\ \ \
i<j\end{array} \right. \ \ \ (W_{ij}=-W_{ji}).\eqno (3.4.8)$$
Recalling ($I_8=\frac{\sqrt{3}}{2}Y$)
\begin{eqnarray*}
&&I^+=\left[ \begin{array}{ccc} 0 & 1 & 0
\\ 0 & 0 & 0 \\ 0 & 0 & 0 \end{array} \right],\ U^+=\left[ \begin{array}{ccc} 0 & 0 & 0
\\ 0 & 0 & 1 \\ 0 & 0 & 0 \end{array} \right],\ V^+=\left[ \begin{array}{ccc} 0 & 0 & 0
\\ 0 & 0 & 0 \\ 1 & 0 & 0 \end{array} \right],\\
\mbox{}\hspace{2cm}&&I^3=\left[ \begin{array}{ccc} 1 & 0 & 0
\\ 0 & -1 & 0 \\ 0 & 0 & 0 \end{array} \right],\ Y=\frac{1}{3}\left[ \begin{array}{ccc} 1 & 0 & 0
\\ 0 & 1 & 0 \\ 0 & 0 & -2 \end{array} \right].\hspace{4.3cm} (3.4.9)
\end{eqnarray*}
We find
\begin{eqnarray*}
J_{\mu}&=&
\{\bar{I}_{\pm},\bar{U}_{\pm},\bar{V}_{\pm},\bar{I}_{3},\bar{I}_{8}\},\\
\bar{I}_{\pm}&=&\sum_i\mu_iI_i^{\pm}\mp2h\sum_{i\neq
j}W_{ij}(I_i^{\pm}I_j^3+\frac{1}{2}U_i^{\mp}V_j^{\mp}),\\
\bar{U}_{\pm} &=& \sum_i\mu_iU_i^{\pm}\pm h\sum_{i\neq
j}W_{ij}[U_i^{\pm}(I_j^3-\frac{3}{2}Y_j)+I_i^{\mp}V_j^{\mp}],\\
\bar{V}_{\pm} &=& \sum_i\mu_iV_i^{\pm}\pm h\sum_{i\neq
j}W_{ij}[V_i^{\pm}(I_j^3+\frac{3}{2}Y_j)+U_i^{\mp}I_j^{\mp}],\\
\bar{I}_3 &=& \sum_i\mu_iI_i^3+h\sum_{i\neq
j}W_{ij}[I_i^+I_j^--\frac{1}{2}(U_i^+U_j^--V_i^+V_j^-)],\\
\mbox{}\hspace{2cm}\bar{I}_8 &=& \sum_i\mu_iY_i+h\sum_{i\neq
j}W_{ij}(U_i^+U_j^--V_j^+V_j^-),\hspace{4.4cm} (3.4.10)
\end{eqnarray*}
where $\mu_i$ and $h$ (not Planck constant) are arbitrary
parameters. Notice again that the simplest choice of $W_{ij}$ is
given by equation (3.4.8).

When $i=1,2$, $Y(SU(2))$ makes transition between spin singlet and
triplet. Now $Y(SU(3))$ transits $SU(3)$ singlet and Octet. For
instance, setting
\begin{eqnarray*}
\hspace{1cm} &&|\ \pi^-\rangle = |d\bar{u}\rangle,\;\;
|\pi^0\rangle=\frac{1}{\sqrt{2}}(|u\bar{u}\rangle-|d\bar{d}\rangle),|K^-\rangle
=|d\bar{u}\rangle,
\ \ |K^0\rangle=|d\bar{s}\rangle,\\
&&|\ \eta^0\rangle
=\frac{1}{\sqrt(6)}(-|u\bar{u}\rangle-|d\bar{d}\rangle+2|s\bar{s}\rangle),
|\ \eta^{0'}\rangle
=\frac{1}{\sqrt(3)}(|u\bar{u}\rangle+|d\bar{d}\rangle+|s\bar{s}\rangle).\hspace{0.7cm}
(3.4.11)
\end{eqnarray*}
Special interest is the following. When
$$\mu_1-\mu_2=-3h,\ f=-\frac{1}{2}(\mu_1-\mu_2),\eqno (3.4.12)$$
by acting the Yangian operators on the Octet of $SU(3)$, we obtain
(see Figure 1)
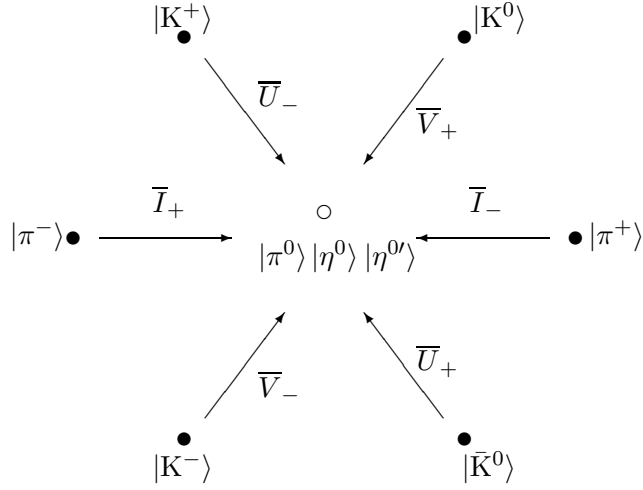
\begin{figure}[ht]

\begin{picture}(440,220)(-160,0)
\put(75,140){\circle{5}} \put(50,120){$| \pi^0 \rangle$ }
\put(70,120){$| \eta^0
\rangle$}\put(90,120){$|\eta^{0\prime}\rangle$}
\put(160,130){\vector(-1,0){50}} \put(170,130){\circle*{5}} \put(175,128){$|\pi^+ \rangle$ }\\
\put(-10,130){\vector(1,0){50}} \put(-20,130){\circle*{5}} \put(-44,128){$|\pi^- \rangle$ }\\
\put(120,198){\vector(-3,-4){30}} \put(128,206){\circle*{5}} \put(131,210){$|$K$^0$$\rangle$ }\\
\put(30,198){\vector(3,-4){30}} \put(22,206){\circle*{5}} \put(10,210){$|$K$^+$$\rangle$ }\\
\put(30,62){\vector(3,4){30}}\put(22,54){\circle*{5}} \put(10,40){$|$K$^-$$\rangle$ }\\
\put(120,62){\vector(-3,4){30}} \put(128,54){\circle*{5}} \put(127,40){$|$\={K}$^0$$\rangle$ }\\
\put(130,140) {$\overline I_-$} \put(10, 140) {$\overline I_+$}
\put(110,170) {$\overline V_+$} \put(50, 180) {$\overline U_-$}
\put(110,80){$\overline U_+$}  \put(50, 70) {$\overline V_-$}
\end{picture}
\caption{Representation of $SU(3)$}
\end{figure}
\begin{eqnarray*}
\hspace{1cm}&&\overline{I}_-|\pi^+>=\frac{1}{\sqrt{6}}(\mu_1-\mu_2)|\eta^0>+
\frac{1}{\sqrt{2}}(\mu_1+\mu_2)|\pi^0>-\frac{1}{\sqrt{3}}(\mu_1-\mu_2+3h)
|\eta^{0'}>,\\
&&\overline{U}_+|\overline{K}^0>=\frac{1}{\sqrt{6}}(\mu_1+2\mu_2)|\eta^0>+
\frac{1}{\sqrt{2}}\mu_1|\pi^0>-\frac{1}{\sqrt{3}}(\mu_1-\mu_2+3h)
|\eta^{0'}>,\\
&&\overline{U}_-|K^0>=\frac{1}{\sqrt{6}}(2\mu_1+\mu_2)|\eta^0>+
\frac{1}{\sqrt{2}}\mu_2|\pi^0>+\frac{1}{\sqrt{3}}(\mu_1-\mu_2+3h)
|\eta^{0'}>,\\
&&\overline{V}_+|K^+>=\frac{1}{\sqrt{6}}(2\mu_1+\mu_2)|\eta^0>-
\frac{1}{\sqrt{2}}\mu_2|\pi^0>+\frac{1}{\sqrt{3}}(\mu_1-\mu_2+3h)
|\eta^{0'}>,\\
&&\overline{V}_-|K^->=-\frac{1}{\sqrt{6}}(\mu_1+2\mu_2)|\eta^0>+
\frac{1}{\sqrt{2}}\mu_1|\pi^0>+\frac{1}{\sqrt{3}}(\mu_1-\mu_2+3h)
|\eta^{0'}>,\\
&&\overline{I}_3|\pi^0>=-\frac{1}{2\sqrt{3}}(\mu_1-\mu_2)|\eta^0>+
\frac{1}{\sqrt{6}}(\mu_1-\mu_2+3h) |\eta^{0'}>,\\
&&\overline{I}_8|\eta^0>=-\frac{1}{3}(\mu_1-\mu_2)|\eta^0>-
\frac{\sqrt{2}}{3}(\mu_1-\mu_2+3h) |\eta^{0'}>,\hspace{2.8cm}
(3.4.13)
\end{eqnarray*}
i.e.,
\begin{eqnarray*}
\hspace{1.5cm}&&(\overline{I}_\pm+fI_\pm)|\eta^{0'}>=\pm2\sqrt{3}h|\pi^{\pm}>,\
(\overline{U}_++fU_+)|\eta^{0'}> = -2\sqrt{3}h|K^0>,\\
&&(\overline{U}_-+fU_-)|\eta^{0'}>=2\sqrt{3}h|\overline{K}^0>,\
(\overline{V}_\pm+fV_\pm)|\eta^{0'}>=-2\sqrt{3}h|K^{\mp}>,\\
&&(\overline{I}_3+fI_3)|\eta^{0'}>=-\sqrt{6}h|\pi^0>,\
(\overline{I}_8+fI_8)|\eta^{0'}>=2\sqrt{2}h|\eta^0>,\hspace{1.5cm}
(3.4.14)
\end{eqnarray*}
and
\begin{eqnarray*}
\hspace{1cm}&&(\overline{I}_\pm+fI_\pm)|\pi^{\mp}>=\pm\sqrt{\frac{3}{2}}h|\eta^0>,\\
&&(\overline{U}_++fU_+)|K^0>=-\frac{\sqrt{3}}{2\sqrt{2}}h(\sqrt{3}|\pi^0>-|\eta^0>),\\
&&(\overline{U}_-+fU_-)|K^0>=\frac{\sqrt{3}}{2\sqrt{2}}h(\sqrt{3}|\pi^0>-|\eta^0>),\\
&&(\overline{V}_\pm+fV_\pm)|K^{\pm}>=-\frac{\sqrt{3}}{2\sqrt{2}}h(\sqrt{3}|\pi^0>+|\eta^0>),\\
\mbox{}\hspace{0.8cm}&&(\overline{I}_3+fI_3)|\pi^0>=\sqrt{\frac{3}{2}}h|\eta^0>,\;
(\overline{I}_8+fI_8)|\eta^0>=\sqrt{3}h|\eta^0>.\hspace{2.5cm}
(3.4.15)
\end{eqnarray*}
The Yangian operators play the role to transit the Octet states to
the singlet state of $SU(3)$.

Whereas, if
$$\mu_1-\mu_2=3h,\ f=-\frac{1}{2}(\mu_1+\mu_2),\eqno (3.4.16)$$
with the notations
$$(\overline{A}^{(2)}+fA^{(1)})|\eta^{0'}>=0,\ A=I_{\alpha},\
(\alpha=\pm,3,8),\ U_{\pm},\ V_{\pm},\eqno (3.4.17)$$ we have
\begin{eqnarray*}
\hspace{2cm}(\overline{I}_\pm+fI_\pm)|\pi^{\mp}>&=&\mp\sqrt{\frac{3}{2}}h|\eta^0>\pm2\sqrt{3}h|\eta^{0'}>,\\
(\overline{U}_++fU_+)|\overline{K}^0>&=&\frac{\sqrt{3}}{2\sqrt{2}}h(\sqrt{3}|\pi^0>-|\eta^0>)-2\sqrt{3}h|\eta^{0'}>,\\
(\overline{U}_-+fU_-)|K^0>&=&-\frac{\sqrt{3}}{2\sqrt{2}}h(\sqrt{3}|\pi^0>-|\eta^0>)+2\sqrt{3}h|\eta^{0'}>,\\
(\overline{V}_\pm+fV_\pm)|K^{\pm}>&=&\frac{\sqrt{3}}{2\sqrt{2}}h(\sqrt{3}|\pi^0>+|\eta^0>)+2\sqrt{3}h|\eta^{0'}>,\\
(\overline{I}_3+fI_3)|\pi^0>&=&-\frac{\sqrt{3}}{2}h|\eta^0>+\sqrt{6}h|\eta^{0'}>,\\
\hspace{1cm}(\overline{I}_8+fI_8)|\eta^0>&=&h|\eta^0>-2\sqrt{2}h|\eta^{0'}>.
\hspace{4.7cm} (3.4.18)
\end{eqnarray*}
Obviously, in this case the Yangian operators make the transition
from the Octet to a ``combined" singlet state of $SU(3)$.


\bigskip

\subsection{{$\bf J ^2$} as a new quantum number}

Because $[{\bf I}^ 2,{\bf J}^2]=0,\ [{\bf I}^2,I_z]=0,[{\bf
J}^2,I_z]=0$, but $[{\bf J}^2, J_z]\ne 0$, we can take $\{{\bf
I}^2,I_z,{\bf J}^2\}$ as a conserved set.

First we consider the case ${\bf S}_1\bigotimes{\bf
S}_2\bigotimes{\bf S}_3$, where $S_1=S_2=S_3=\frac{1}{2}$. We
shall show that instead of 6-j coefficients and Young diagrams,
${\bf J}^2$ can be viewed as a ``collective" quantum number that
describes the ``history" besides ${\bf S}^2$ $({\bf S}={\bf
S}_1+{\bf S}_2+{\bf S}_3)\ $and$\ S_z$.

As representations of Lie algebra $SU(2)$, we have
$$(\frac{1}{2}\bigotimes\frac{1}{2})\bigotimes\frac{1}{2}=(1\bigoplus
0)\bigotimes\frac{1}{2}=\frac{3}{2}\bigoplus\frac{1}{2}\bigoplus\frac{1}{2}'.\eqno
(3.5.1)$$ Noting that $|\frac{1}{2}\rangle$ and
$|\frac{1}{2}'\rangle$ are degenerate regarding the total spin
$\frac{1}{2}$. The usual Lie algebraic base can be easily written
as
\begin{eqnarray*}
\hspace{4cm}&&\phi_{\frac{3}{2},\frac{3}{2}}=|\uparrow\uparrow\uparrow\rangle,\\
&&\phi_{\frac{3}{2},\frac{1}{2}}=\frac{1}{\sqrt{3}}(|\uparrow\uparrow\downarrow\rangle+
|\uparrow\downarrow\uparrow\rangle+|\downarrow\uparrow\uparrow\rangle),\\
&&\phi_{\frac{3}{2},-\frac{1}{2}}=\frac{1}{\sqrt{3}}(|\uparrow\downarrow\downarrow\rangle+
|\downarrow\uparrow\downarrow\rangle+|\downarrow\downarrow\uparrow\rangle),\\
&&\phi_{\frac{3}{2},-\frac{3}{2}}=|\downarrow\downarrow\downarrow\rangle,\hspace{6.8cm}
(3.5.2)
\end{eqnarray*}
and the two degeneracy states with respect to ${\bf S}^2\ $and$\
S_z$ are given by:
\begin{eqnarray*}
&\hspace{4cm}&\phi'_{\frac{1}{2},\frac{1}{2}}=\frac{1}{\sqrt{6}}(|\downarrow\uparrow\uparrow\rangle+
|\uparrow\downarrow\uparrow\rangle-2|\uparrow\uparrow\downarrow\rangle),\\
&&\phi'_{\frac{1}{2},-\frac{1}{2}}=\frac{1}{\sqrt{6}}(|\uparrow\downarrow\downarrow\rangle+
|\downarrow\uparrow\downarrow\rangle-2|\downarrow\downarrow\uparrow\rangle ), \\
&&\phi_{\frac{1}{2},\frac{1}{2}}=\frac{1}{\sqrt{2}}(|\downarrow\uparrow\uparrow\rangle-\uparrow\downarrow\uparrow\rangle),\\
&&\phi_{\frac{1}{2},-\frac{1}{2}}=\frac{1}{\sqrt{2}}(|\uparrow\downarrow\downarrow\rangle-
|\downarrow\uparrow\downarrow\rangle).\hspace{4.8cm} (3.5.3)
\end{eqnarray*}
To distinguish $\phi'$ from $\phi$ we introduce {\bf J}:
$${\bf J}=\sum\limits_{i=1}^3u_i{\bf S}_i+ih\sum\limits_{i<j}^3({\bf
S}_i\times{\bf S}_j),\eqno (3.5.4)$$ and calculate ${\bf J}^2$. It
turns out that \begin{eqnarray*} \hspace{2cm}{\bf
J}^2\phi_{\frac{3}{2},m}&=&[\frac{3}{4}(u_1^2+u_2^2+u_3^2)+\frac{1}{2}(u_1
u_2 + u_2 u_3 + u_1 u_3)-h^2]\Phi_{\frac{3}{2},m};\\
{\bf J}^2 \phi'_{\frac{1}{2},m}& = & [\frac{3}{4} ( u_1^2 + u_2^2
+ u_3^2 ) + \frac{1}{2} u_1 u_2 - u_2 u_3 - u_1 u_3 - \frac{7}{4}
h^2 ]\Phi'_{\frac{1}{2},m}\\
&\mbox{}& -\frac{\sqrt{3}}{2} ( u_1 - u_2 + h ) (
u_3 + h) \Phi_{\frac{1}{2},m};\\
{\bf J}^2 \phi_{\frac{1}{2},m} &=& -\frac{\sqrt{3}}{2}(u_1 - u_2 -
h )( u_3 - h)\Phi'_{\frac{1}{2},m}+[\frac{3}{4} ( u_1 - u_2 )^2\\
&\mbox{}&+ \frac{3}{4} u_3^2 - \frac{3}{4} h^2]
\Phi_{\frac{1}{2},m}. \hspace{7cm}(3.5.5)
\end{eqnarray*}
In order to make the matrix of ${\bf J}^2$ be symmetric (then it
surely can be diagonalized), one should put
$$u_2 = u_1 + u_3.\eqno (3.5.6)$$
The eigenvalues of ${\bf J}^2$ are given by
\begin{eqnarray*}
&&\lambda_{\frac{3}{2}} = 2 u_1^2 + 2 u_3^2 + 3 u_1 u_3 - h^2,\\
\hspace{1.8cm}&&\lambda_{\frac{1}{2}}^\pm = u_1^2 + u_3^2 -
\frac{5}{4} h^2 \pm \frac{1}{2} [ ( 2 u_1^2 - u_3^2 - h^2 )^2 + 3
( u_3^2 - h^2 )^2 ]^{\frac{1}{2}}.\hspace{2cm} (3.5.7)
\end{eqnarray*}
The eigenstates of ${\bf J}^2$ are the rotation of
$\phi'_{\frac{1}{2},m}$ and $\Phi_{\frac{1}{2},m}$:
$$\left( \begin{array}{c} \alpha_{\frac{1}{2},m}^+  \\
\alpha_{\frac{1}{2},m}^-
\end{array} \right) = \left( \begin{array}{cc} \cos\frac{\varphi}{2} & -\sin\frac{\varphi}{2} \\
 \sin\frac{\varphi}{2} & \cos\frac{\varphi}{2} \end{array} \right)
  \left( \begin{array}{c} \phi'_{\frac{1}{2},m} \\ \phi_{\frac{1}{2},m} \end{array} \right),\ \  {\bf J}^2
  \alpha^\pm_{\frac{1}{2}}=\lambda^\pm_{\frac{1}{2}}\alpha^\pm_{\frac{1}{2},m},\eqno (3.5.8)$$
where
$$\sin\varphi=\sqrt{3}( u_3^2 - h^2 ) / \omega,\;\;\omega^2 = ( 2 u_1^2 - u_3^2 - h^2 )^2 + 3
( u_3^2 - h^2 )^2.\eqno (3.5.9)$$ It is worth noting that the
conclusion is independent of the order, say, $( \frac{1}{2}
\bigotimes \frac{1}{2} ) \bigotimes \frac{1}{2}$, $\frac{1}{2}
\bigotimes (\frac{1}{2} \bigotimes \frac{1}{2})$ and the other
way. The difference is only in the value of $\varphi$.

\vskip 0.5cm

The above example can be generalized to ${\bf S}_1 \bigotimes {\bf
S}_2 \bigotimes {\bf L}$ where $S_1=S_2=\frac{1}{2}$ and ${\bf
L}^2=l(l+1)$. As representations of Lie algebra $SU(2)$, we have
\begin{eqnarray*}\hspace{3cm}(\frac{1}{2} \bigotimes \frac{1}{2} ) \bigotimes l = (1
\bigoplus 0 ) \bigotimes l = l + 1\ \ \ \ \ &l& \ \ \ \ \
l-1\\
&l& \hspace{3.6cm} (3.5.10)\end{eqnarray*}

There are no degeneracy for $l\pm1$, but two $l$ states can be
distinguished in terms of ${\bf J}^2$
\begin{eqnarray*}\hspace{1cm}{\bf J}^2 \Phi_{l+1,m} &=& \{ \frac{3}{4} ( u_1^2
+u_2^2 ) + l ( l
+ 1 ) u_3^2 + \frac{1}{2} u_1 u_2 + l ( u_2 u_3 + u_1 u_3) \\
& &- h^2 [
l ( l + 1 ) + \frac{1}{4} ] \} \Phi_{l + 1 , m},\\
{\bf J}^2 \Phi_{l-1,m} &=& \{ \frac{3}{4} ( u_1^2 + u_2^2 ) + l (
l + 1 ) u_3^2 + \frac{1}{2} u_1 u_2 -( l + 1 ) u_1 u_3 - (l + 1)
u_2
u_3 \\
& &- h^2 [ l (l + 1) + \frac{1}{4} ] \} \Phi_{l-1,m},\\
{\bf J}^2 \Phi_{l,m}^1 &=& \{ \frac{3}{4} ( u_1^2 + u_2^2 ) + l(l
+ 1) u_3^2 + \frac{1}{2} u_1 u_2 - u_2 u_3 - u_1 u_3\\
&& - 2 h^2 [l(l + 1) \frac{1}{8} ]\} \Phi_{l,m}^1- \sqrt{l(l + 1)}
(
u_1 - u_2 + h)(u_3 + h)\Phi_{l,m}^2,\\
{\bf J}^2 \Phi_{l,m}^2 &=& - \sqrt{l(l + 1)} (u_1 - u_2 - h)(u_3 -
h) \Phi_{l,m}^1 \\&&+ [\frac{3}{4} (u_1 - u_2)^2 + l(l + 1) u_3^2
- \frac{3}{4}] \Phi_{l,m}^2.\hspace{4.5cm} (3.5.11)
\end{eqnarray*}
Again in order to guarantee the symmetric form of the matrix we
put $$u_2=u_1+u_3,\eqno (3.5.12)$$ then the eigenvalues and
eigenstates of ${\bf J}^2$ are given by
$$\lambda^\pm_l=u^2_1+[l(l+1)+\frac{1}{4}]u^2_3-h^2[l(l+1)+\frac{1}{2}]\pm\frac{1}{2}\sqrt{P},\eqno (3.5.13)$$
$${\left(\matrix{\alpha^+_{l,m} \cr
\alpha^-_{l,m}}\right)}={\left(\matrix{\cos\frac{\varphi}{2} &
-\sin\frac{\varphi}{2} \cr \sin\frac{\varphi}{2} &
\cos\frac{\varphi}{2}}\right)}{\left(\matrix{\Phi^1_{l,m} \cr
\Phi^2_{l,m}}\right)},\eqno (3.5.14)$$ where
$$\omega^2=P=[2u_1^2-u^2_3-h^2(2l(l+1)-\frac{1}{2})]^2+4l(l+1)(u^2_3-h^2)^2,\eqno
(3.5.15)$$
$$\sin\varphi=\frac{2\sqrt{l(l+1)}}{\omega}(u^2_3-h^2).\eqno
(3.5.16)$$

\vskip 0.5cm

As a simple example, we consider the spin structure of rare gas
$$H=-a{\bf L}\cdot{\bf S}_1-b{\bf S}_1\cdot{\bf S}_2,\;\;
(\lambda=\frac{b}{a}).\eqno (3.5.17)$$ It describes the
interaction of spin ${\bf S}_1$ of an electron exited from
$l$-shell and the left hole ${\bf S}_2$.
\begin{eqnarray*}
\hspace{2.4cm}&&
H\Phi_{l+1,m}=-\frac{1}{2}(al+\frac{1}{2}b)\Phi_{l+1,m},\\
&&
H\Phi_{l-1,m}=\frac{1}{2}[(l+1)a-\frac{1}{2}b]\Phi_{l-1,m},\\
&& H{\left[\matrix{\Phi^\pm_{l,m} \cr
\Phi^2_{l,m}}\right]}=\frac{1}{2}{\left[\matrix{(a-\frac{1}{2}b) &
a\sqrt{l(l+1)} \cr a\sqrt{l(l+1)} &
\frac{3}{2}b}\right]}{\left[\matrix{\Phi^1_{l,m} \cr
\Phi^2_{l,m}}\right]}.\hspace{3cm} (3.5.18)
\end{eqnarray*}
The eigenstates of $H$ associated to $l,m$ are
$${\left(\matrix{\alpha^+_{l,m} \cr
\alpha^-_{l,m}}\right)}={\left(\matrix{\cos\frac{\varphi}{2} &
-\sin\frac{\varphi}{2} \cr \sin\frac{\varphi}{2} &
\cos\frac{\varphi}{2}}\right)}{\left(\matrix{\Phi^1_{l,m} \cr
\Phi^2_{l,m}}\right)}.\eqno (3.5.19)$$ where
$$\sin\varphi=\frac{\sqrt{l(l+1)}}{\omega},\;\;\omega^2=(\frac{1}{2}-\lambda)^2+l(l+1),
\lambda=\frac{b}{a}. \eqno (3.5.20)$$ The eigenvalues are
\begin{eqnarray*}
\hspace{2cm}&&\lambda_{l+1}=-\frac{1}{2}(la+\frac{b}{2}),\ \
\lambda_{l-1}=\frac{1}{2}[(l+1)a-\frac{b}{2}];\\
&&\lambda^\pm_l=\frac{1}{4}(a+b)\pm\frac{1}{2}[l(l+1)a^2+(\frac{a}{2}-b)^2]^{\frac{1}{2}}.\hspace{4.4cm}
(3.5.21)
\end{eqnarray*}
The rotation should be made in such a way that
$$[H,{\bf J}^2]=0\eqno (3.5.22)$$
which is satisfied if the matrix ${\bf J}^2$ is symmetric, i.e.,
$$\gamma = \frac{\{2u^2_1-2h^2[l(l+1)+\frac{1}{4}]\}}{(u^2_3-h^2)}
= 2(1-\lambda).\eqno (3.5.23)$$ Therefore, the parameter $\gamma$
in $Y(SU(2))$ determines the rotation angle $\varphi$. It is
reasonable to think that the appearance of ``rotation" of
degenerate states is closely related to the ``quantum number" of
${\bf J}^2$. Transition between $\alpha^+_{l,m}$ and
$\alpha^-_{l,m}$ $(l=1)$ can be made by $J_3$. Because there are
two independent parameters $u_1$ and $u_3$ in ${\bf J}$, one can
choose a suitable relation between $u_3$ and $\lambda=\frac{b}{a}$
such that
$$J_3\alpha^+\sim\alpha^-,\eqno (3.5.24)$$
i.e., the transition between two degenerate states in Lie-algebra
is made trough $J_3$ operator, because of
$$[{\bf J}^2,J_3]\neq 0.\eqno (3.5.25)$$

\bigskip

\subsection{Happer degeneracy}

In the experiment for ${}^{87}R_b$ molecular there appears new
degeneracy ([41]) at the special $\pm B_0$ (magnetic field), i.e.,
the Zeeman effect disappears at $\pm B_0$. The model Hamiltonian
reads ([42]) ($x$ is scaled magnetic field)
$$H={\bf K}\cdot{\bf S}+x(k+\frac{1}{2})S_z,\eqno (3.6.1)$$
where ${\bf K}$ is angular momentum and ${\bf K}^2=K(K+1)$. It
only occurs for spin $S=1$. It turns out that when $x=\pm 1$ there
appears the curious degeneracy, that is, there is a set of
eigenstates corresponding to
$$E=-\frac{1}{2}.\eqno (3.6.2)$$
The conserved set is $\{{\bf K}^2,
G_z=K_z+S_z\}$. For ${\bf G}={\bf K}+{\bf S}$ we have $G=k\pm 1,
k$. The eigenstates are specified in terms of three families: $T,
B$ and $D$. Only D-set possesses the degeneracy.

Happer gives, for example,the eigenstates for $x=\pm1$ ([42]):
$$\matrix{x=+1 & \mbox{}\hspace{1.5cm}\mbox{}&
H\alpha_{Dm}=(-\frac{1}{2})\alpha_{Dm}\cr x=-1 & &
H\beta_{Dm}=(-\frac{1}{2})\beta_{Dm}},\eqno (3.6.3)$$ and shows
that \begin{eqnarray*}
\hspace{1cm}\alpha_{Dm}&=&[2(K+\frac{1}{2})(K+m+\frac{1}{2})]^{-\frac{1}{2}}\{-[\frac{(K-m+1)(K+m+1)}{2}]^{\frac{1}{2}}\alpha_1\\
&\mbox{}&+[(K+m)(K+m+1)]^{\frac{1}{2}}\alpha_2+[\frac{(K-m)(K+m)}{2}]^{\frac{1}{2}}\alpha_3\};\hspace{2.2cm} (3.6.4)\\
\beta_{Dm}&=&[2(K+\frac{1}{2})(K-m+\frac{1}{2})]^{-\frac{1}{2}}\{[\frac{(K-m)(K+m)}{2}]^{\frac{1}{2}}\alpha_1\\
&\mbox{}&+[(K-m)(K-m+1)]^{\frac{1}{2}}\alpha_2-[\frac{(K-m+1)(K+m+1)}{2}]^{\frac{1}{2}}\alpha_3\},\hspace{0.8cm}
(3.6.5)
\end{eqnarray*}
where $\alpha_1=e_1\otimes e_{m-1}$, $\alpha_2=e_0\otimes e_m$ and
$\alpha_3=e_{-1}\otimes e_{m+1}$.

It is natural to ask what is the transition operator between
$\alpha_{Dm}$ and $\beta_{Dm}$? The answer is Yangian operator. In
fact, introducing
$$J_\pm=aS_++bK_-\pm(s_\pm
K_z-s_zK_\pm),\eqno (3.6.6)$$ we find that by choosing
$a=-\frac{k+1}{2},b=0$, we have
$$\beta_{Dm}\stackrel{J_+}{\longrightarrow}\lambda_1(m)\alpha_{D
m+1}\;\;{\rm
and}\;\;\alpha_{Dm}\stackrel{J_-}{\longrightarrow}\lambda_2(m)\beta_{D
m-1};\eqno (3.6.7)$$ and by choosing $a=\frac{k}{2},b=0$, we have
$$\beta_{Dm}\stackrel{J_-}{\longrightarrow}\lambda'_1(m)\alpha_{D
m-1}\;\;{\rm
and}\;\;\alpha_{Dm}\stackrel{J_+}{\longrightarrow}\lambda'_2(m)\beta_{D
m+1}.\eqno (3.6.8)$$

The Yangian makes the transition between the states with $B$ and
$-B$, which here is only for $S=1$. The reason is that for $S=1$
there are two independent coefficients in the combination of
$\alpha_1,\alpha_2$ and $\alpha_3$ and there are two free
parameters in ${\bf J}$. Hence the number of equations are equal
to those of free parameters ($a$ and $b$), so we can find a
solution. The numerical computation shows that only $S=1$ gives
rise to the new degeneracy ([42]) that prefers the Yangian
operation ([43]).

\bigskip

\subsection{New degeneracy of extended Breit-Rabi
Hamiltonian}

As was shown in the Happer's model ($H={\bf K}\cdot {\bf
S}+x(k+\frac{1}{2})S_3$) there appeared new degeneracy for $S=1$.
It has been pointed out that the above degeneracy with respect to
Zeeman effect cannot appear for spin=$\frac{1}{2}$. Actually, in
this case it yields for $S=\frac{1}{2}$ ([42]),
$$E=-\frac{1}{4}-\omega_mS_3,\eqno (3.7.1)$$
where
$$\omega_m^2=[(1+x^2)(k+\frac{1}{2})+2xm](k+\frac{1}{2}).\eqno (3.7.2)$$
Therefore if the Happer's type of degeneracy can occurs, there
should be $\omega_m=0$ that means
$$x_0=-\frac{m}{k}\pm i
\sqrt{1-\frac{m^2}{k^2}}\;\;(k=K+\frac{1}{2}),\eqno (3.7.3)$$
i.e., the magnetic field should be complex.

However, the situation will be completely different, if a third
spin is involved. For simplicity we assume
$S_1=S_2=S_3=\frac{1}{2}$ in the Hamiltonian:
$$H=-(a{\bf S}_2+b{\bf S}_3)\cdot {\bf
S}_1+x\sqrt{ab}S_1^z,\lambda=b/a,\eqno (3.7.4)$$ then besides two
non-degenerate states, there appears the degenerate family:
$$H\alpha_{D,\pm\frac{1}{2}}^\pm=-(\frac{a+b}{4})\alpha_{D,\pm\frac{1}{2}}^\pm,\;\;{\rm
for}\;\;x=\pm 1,\eqno (3.7.5)$$ where
$$\alpha_{D,+\frac{1}{2}}^\pm=-\sqrt{2}\lambda |\uparrow\uparrow\downarrow>\pm \sqrt{\lambda}|\uparrow\downarrow\uparrow
+(1\pm\sqrt{\lambda})|\downarrow\uparrow\uparrow>;\hspace{1cm}\mbox{}\eqno
(3.7.6)$$
$$\alpha_{D,-\frac{1}{2}}^\pm=-\sqrt{2}\lambda |\downarrow\downarrow\uparrow>\mp \sqrt{\lambda}|\downarrow\uparrow\downarrow
+(1\mp\sqrt{\lambda})|\uparrow\downarrow\downarrow>.\eqno
(3.7.7)$$ The expecting value of $S_1^z$ are
$$<\alpha_{D,\pm\frac{1}{2}}^+|S_1^z|\alpha_{D,\pm\frac{1}{2}}^+>\sim
\sqrt{\lambda}\; (x=1);\eqno (3.7.8)$$
$$<\alpha_{D,\pm\frac{1}{2}}^-|S_1^z|\alpha_{D,\pm\frac{1}{2}}^->\sim
-\sqrt{\lambda}\; (x=-1).\eqno (3.7.9)$$ namely, at the special
magnetic field ($x=\pm 1$) the observed $<S_1^z>$ still opposite
to each other for $x=\pm 1$, but without the usual Zeeman split.

The reason of the appearance of the new degeneracy is obvious. The
two spins ${\bf S}_2$ and ${\bf S}_3$ here play the role of $S=1$
in comparison with Happer model.

\bigskip

\subsection{Super Yang-Mills ($N=4$)-Lipatov model and $Y(SO(6))$}

Beisert et al([44-45]), Dolan-Nappi-Witten (DNW,[34]) and other
authors ([46-47]) proposed to take the quantum correction of the
dilatation operator $\delta D$ ($D\in SO(4,2)$ is a subalgebra of
$PSU(2,2|4))$ as Hamiltonian for supper Yang-Mills $ (N=4)$:
$$H=\sum_\alpha H_{\alpha\alpha+1},\eqno (3.8.1)$$
$$ H_{\alpha\alpha+1}=2\sum_jh(j)P^j_{\alpha\alpha+1},\ \
h(j)=\sum^j_{k=1}\frac{1}{k}, h(0)=1.\eqno (3.8.2)$$ where $P^j$
is projector for the weight $j$ of $SU(2)$ and $\alpha$ stands for
``lattice'' index. DNW showed that ([34])
$$[H,Y(SO(6))]=0.\eqno (3.8.3)$$ It turns out that the Hamiltonian
$H$ is nothing but Lipatov model ([48]) which was related to the
Yang-Baxter form by Lipatov ([49]), Faddeev and Korchemsky ([50]).

Based on Tarasov, Takhtajan and Faddeev([51]) the
$\breve{R}$-matrix associated with any spin $S$ reads
$$\breve{R}(u)=\frac{\Gamma(u-s)\Gamma(u+2s+1)}{\Gamma(u-\hat{J})\Gamma(u+\hat{J}+1)},\eqno
(3.8.4)$$ where $u$ is spectrum parameter and $s$ the spin
(arbitrary). The trigonometric Yang-Baxteri-zation ([52]) gives
$$\breve{R}(u)=\sum_{j=0}\rho_j(x)P_j(q)\;\; (x=e^{iu}),\eqno
(3.8.5)$$ where $P_j(q)$ is the $q$-deformed product with weight
$j$. Taking the rational limit ([9],[36]) we have
$$\rho_j\Rightarrow\frac{\Gamma(u)\Gamma(u+1)}{\Gamma(u-j)\Gamma(u+j+1)},\
\ P_j(q)\Rightarrow P_j.\eqno (3.8.6)$$ The Hamiltonian for the
lattices $\alpha$ and $\alpha+1$ $$ H_{\alpha\alpha+1}=I_1\times
I_2\times\cdots\times
I_{\alpha-1}\times\frac{d}{du}\breve{R}(u)|_{u=0}[\breve{R}(0)]^{-1}\times
I_{\alpha+2}\times\cdots\eqno (3.8.7)$$
 is then
$$H=\sum_\alpha H_{\alpha\alpha+1}\eqno (3.8.8)$$
where  \begin{eqnarray*}
\hspace{1cm}H_{\alpha\alpha+1}&=&\{-\psi(-\hat{J}_{\alpha\alpha+1})-\psi(\hat{J}_{\alpha\alpha+1}+1)+\psi(1+2s)+\psi(1-2s)-
\frac{1}{2s}\}|_{s=0}\\
&=&\sum_j\{-\psi(-j)-\psi(j+1)+2\psi(1)-\lim_{x\rightarrow0}\frac{1}{x}\}P^j_{\alpha\alpha+1}.\hspace{3.2cm}
(3.8.9) \end{eqnarray*} It describes the QCD correction to the
parton model shown by Lipatov ([48-49]). The diagonalization of
Lipatov model has probably been achieved by de Vega and Lipatov
([53-54]). Noting that the $j$ indicates the block in the
reducible block-diagonal form.

Using
\begin{eqnarray*}
\hspace{4cm}&& \psi(x+1)=\psi(x)+\frac{1}{x},\\
&& \psi(x+n)=\psi(x)+\sum^{n-1}_{k=0}\frac{1}{x+k},\\
&& \psi(1)=-c,\hspace{7.5cm} (3.8.10)
 \end{eqnarray*}
and hence
\begin{eqnarray*}
\hspace{3cm}&&
\psi(j+1)=\psi(1)+\sum^j_{k=1}\frac{1}{k}=\psi(1)+h(j)\\
&&
\psi(-j)=\psi(1)+h(j)-\lim_{x\rightarrow0}\frac{1}{x}.\hspace{5.5cm}
(3.8.11)
 \end{eqnarray*}
We obtain
$$H_{\alpha,\alpha+1}=(-2)\sum_jh(j)P^j_{\alpha\alpha+1}.\eqno
(3.8.12)$$ Separating the finite part from the infinity the $H$ is
nothing but the $\delta D$ derived in super Yang-Mills $(N=4)$
with the approximation. Of course, the derivation of $\delta D$
based on super Yang-Mills ($N=4$) explores much larger symmetry
than Lipatov model. Therefore, DNW's result shows that the
Lipatov's model possesses $Y(SO(6))$ symmetry.

To obtain $Y(SO(6))$ in terms of RTT relation we start from the
rational solution of $\breve{R}$-matrix whose general form for
$O(N)$ was firstly by Zamolodchikov and Zamolodchikov ([35]) and
extended through rational limit of trigonometric
Yang-Baxterization ([36]):
$$\breve{R}=u[u-\frac{1}{2}(N-2)a]P+\alpha
uA_N+[-u\alpha+\frac{\alpha^2}{2}(N-2)]I.\eqno (3.8.13)$$ where
$u$ is spectrum parameter and $\alpha$ a free parameter allowed by
YBE. Here we adopt the convention of Jimbo: $$
P^{ab}_{cd}=\delta^a_d\delta^b_c,\;\;(A_N)^{ab}_{cd}=\delta^{a,-b}\delta_{c,-d}
 \eqno (3.8.14)$$
where
$$a,b,c,d=[-(\frac{N-1}{2}),-(\frac{N-1}{2})+1,\cdots,(\frac{N-1}{2})]\eqno
(3.8.15)$$ and $N=2n+1$ for $B_n$ and $N=2n$ for $C_n$, $D_n$.

The R-matrix is given by
$$R=\breve{R}P=u(u-2\alpha)I+u(2u-\alpha)P+2u\alpha A_N,\eqno
(3.8.16)$$ that coincides with Zamolodchikov's $S$-matrix (up to
an over all factor considering the CDD poles) with $\alpha=1$ and
$u=\frac{\theta}{i\lambda}$. Actually, Zamolodchikov's $S$-matrix
is universal, i.e., model independent.
\begin{eqnarray*}
\hspace{3cm}S(\theta)=R(u)&=&
Q^\pm(u)u(u-2)[I+\frac{\sigma_3}{\sigma_2}P+\frac{\sigma_1}{\sigma_2}A_N]\nonumber\\
&=& Q^\pm(u)u(u-2)[I-\frac{1}{u}P+\frac{2}{u-2}A_N],\nonumber\\
Q^\pm(u) &=&
\frac{\Gamma(\pm\frac{\lambda}{2\pi}-i\frac{\theta}{2\pi})\Gamma(\frac{1}{2}-i\frac{\theta}{2\pi})}{
\Gamma(\frac{1}{2}\pm\frac{\lambda}{2\pi}-i\frac{\theta}{2\pi})\Gamma(-i\frac{\theta}{2\pi})}
 \hspace{4cm} (3.8.17)\end{eqnarray*}
where $\lambda=\frac{2\pi}{N-2}$, $\theta=i\lambda u$. The
spectrum parameter $u$ is one-dimensional, but $u$ can be taken to
be the cut-off in 4-dimensional quantum field theory, for example
$$
u\sim\ln\Lambda^2,\eqno (3.8.18)$$ where $\Lambda^2$ is Lorentz
invariant, i.e., scalar. This is the reason why asymptotic
behavior of quantum field theory model may be related to
Yang-Baxter system. The Bethe Ansatz for $S(\theta)$ with $SO(6)$
was discussed by Minahan and Zarembo ([46]).

For given $\breve{R}(u)$ one can easily obtain Hamiltonian by
$$H=[\frac{\partial\breve{R}(u)}{\partial
u}\breve{R}(u)]|_{u=0},\eqno (3.8.19)$$ for $O(N)$.

However, the essential connection between Lipatov model and
$SO(6)$-RTT formulation is still missing.

\bigskip

\section{Remarks}

Although there has been certain progress of Yangian's application
in physics, there are still open questions:

(1) How can the Yangian representations  help to solve physical
models, in particular, in strong correlation models?

(2) Direct evidences of Yangian in the real physics.

(3) What is the geometric meaning of Yangian?

\bigskip

\section*{Acknowledgement}

We thank Professors F. Dyson and W. Happer for encouragement and
enlighten discussions. This work is in part supported by NSF of
China.

\end{document}